**ADVANCED REVIEW** OPEN ACCESS

# Fragility Modeling of Power Grid Infrastructure for Addressing Climate Change Risks and Adaptation

George Karagiannakis[1] | Mathaios Panteli[2] | Sotirios Argyroudis[3,4]

[1]Department of Architecture, Built Environment and Construction Engineering, Politecnico di Milano, Milan, Italy | [2]Department of Electrical and Computer Engineering, University of Cyprus, Nicosia, Cyprus | [3]Department of Civil and Environmental Engineering, Brunel University of London, London, Middlesex, UK | [4]www.MetaInfrastructure.org, London, UK

**Correspondence:** Sotirios Argyroudis (sotirios.argyroudis@brunel.ac.uk)

**Received:** 23 October 2023 | **Revised:** 29 October 2024 | **Accepted:** 8 November 2024

**Domain Editor:** Timothy R. Carter | **Editor-in-Chief:** Daniel Friess

**Funding:** The second author received funding from the European Union HORIZON-MSCA-2021-SE-01 (grant agreement no. 101086413) ReCharged—Climate-aware Resilience for Sustainable Critical and interdependent Infrastructure Systems enhanced by emerging Digital Technologies. The third author received funding from the UK Research and Innovation (UKRI) under the UK government's Horizon Europe funding guarantee (Ref. EP/X037665/1). This is the funding guarantee for the European Union HORIZON-MSCA-2021-SE-01 (grant agreement no. 101086413) ReCharged project.

## ABSTRACT

The resilience of electric power grids is threatened by natural hazards. Climate-related hazards are becoming more frequent and intense due to climate change. Statistical analyses clearly demonstrate a rise in the number of incidents (power failures) and their consequences in recent years. Therefore, it is of utmost importance to understand and quantify the resilience of the infrastructure to external stressors, which is essential for developing efficient climate change adaptation strategies. To accomplish this, robust fragility and other vulnerability models are necessary. These models are employed to assess the level of asset damage and to quantify losses for given hazard intensity measures. In this context, a comprehensive literature review is carried out to shed light on existing fragility models specific to the transmission network, distribution network, and substations. The review is organized into three main sections: damage assessment, fragility curves, and recommendations for climate change adaptation. The first section provides a comprehensive review of past incidents, their causes, and failure modes. The second section reviews analytical and empirical fragility models, emphasizing the need for further research on compound and non-compound hazards, especially windstorms, floods, lightning, and wildfires. Finally, the third section examines risk mitigation and adaptation strategies in the context of climate change. This review aims to improve the understanding of approaches to enhance the resilience of power grid assets in the face of climate change. These insights are valuable to various stakeholders, including risk analysts and policymakers, who are involved in risk modeling and developing adaptation strategies.

## 1 | Introduction

### 1.1 | Background and Motivation

The electrical power grid is the backbone of modern societies, as most services such as water, transportation, and communication and critical facilities (e.g., health-care and gas) rely heavily on electric power. Ensuring uninterrupted electricity is essential for public safety and business continuity. Every year, however, the functionality of the electrical grid is threatened by weather and other natural hazards including windstorms, wildfires, and earthquakes, among others. In Europe, between 2018 and 2021, it is estimated that 18%–22% of power disruptions were triggered by climate hazards (ENTSO-E 2022). However, the European Commission believes this figure could be as high as 33%, due to under reporting (EC 2018). The World Bank (Hallegatte, Rentschler, and Rozenberg 2019) raised the estimate to 37%, noting that outages from natural hazards lasted around four times longer than those from non-natural causes. These percentages are consistent with global data (Zhaohong et al. 2017) and in some regions can be even higher (ENTSO-E 2013). Climate-induced incidents were evaluated to be the second most frequent cause of power interruptions (ENTSO-E 2022). Recent examples include the 2021 floods in Central Europe or the 2019 heat wave in California, USA, which triggered widespread outages that lasted for over a month and resulted in repair costs of millions and liability costs of billions of dollars (City-Journal 2019; Koks et al. 2022). The excessive losses stem from the strong interdependencies between the power grid and other infrastructure systems and communities (Roege et al. 2014).

Additionally, there is evidence indicating a continuous rise in the number of incidents (i.e., power failures) and their ramifications, primarily driven by climate change effects. Tavares da Costa, Krausmann, and Hadjisavvas (2023) estimated that the average interruption frequency due to exceptional weather events more than doubled in 27 EU + UK countries between 2004 and 2016. Also, the North American Electric Reliability Corporation (NERC) reported that 69 out of 70 large disruptions in power transmission in the United States between 2016 and 2021 were induced by extreme weather conditions (NERC 2022). Furthermore, in 2023, the Intergovernmental Panel on Climate Change stated that every additional increment of global warming will result in changes in extremes, which may (or may not) increase in intensity and frequency, depending on the type of hazard and region (IPCC 2023).

One important contribution to an improved understanding of power grid vulnerabilities is to enhance the robustness of fragility models. Fragility curves (FCs) are simple to understand and useful tools for vulnerability (or loss) assessment of critical infrastructure (Argyroudis et al. 2019; Nirandjan et al. 2024). In conjunction with recovery curves, vulnerability models are an essential component of power grid resilience quantification and assessment (Panteli and Mancarella 2017). Other vulnerability models can be used to estimate resilience. For example, HAZUS methodology proposes FCs only for the earthquake hazard (HAZUS 2022a), whereas damage-to-loss assessment models are described for other types of hazards (HAZUS 2022b, 2022c). Nevertheless, it is worth acknowledging that, by their own admission, these models are likely to be subject to significant bias, considering that they are often based on expert judgment, combined with empirical data. Their application to extreme events is limited due to the scarcity of damage data for such hazard events and the variations in power grid assets across different geographic regions. In addition to fragility and damage-to-loss models, indicators—such as exposure, availability of resources for recovery, and competence of authorities—are also used for vulnerability analysis (Hofmann, Kjølle, and Gjerde 2013; Sperstad, Kjølle, and Gjerde 2020). However, they cannot be used to evaluate losses. These aspects make analytical or hybrid FCs a reliable tool for uncertainty reduction in vulnerability estimation and losses quantification, as it has been demonstrated for other types of assets, for example, bridges (Argyroudis and Mitoulis 2021; Maroni et al. 2022) or chemical plants (Karagiannakis et al. 2022).

Considering their importance, the existing literature on fragility models for power grid assets against natural hazards remains surprisingly scarce, especially when taking into account climatic projections and multiple hazards. Dumas, Kc, and Cunliff (2019) and Kabre and Weimar (2022) carried out a review of power grid vulnerability, which are two out of very few in the literature, and corroborated that the majority of the available models are empirical and single hazards. Nevertheless, these reports primarily provided summaries rather than a critical examination of the characteristics of assets related to failure modes and climate hazards, including their associated intensity parameters. Also, Serrano-Fontova et al. (2023) conducted a comprehensive review of FCs for power grid assets against natural hazards. However, important considerations in facilitating the selection of existing, or the development of new FCs, were not examined thoroughly. These considerations primarily relate to the identification of failure modes and relevant structural parameters, the diverse typologies of assets, modeling uncertainty due to climate change scenarios and, implementation of adaptation strategies. Furthermore, the existing fragility models for electrical grids focus mostly on wind hazard, albeit heavy snow, lightning, floods, and wildfires can also be damaging stressors (NERC 2022).

In light of the urgency to implement improved disaster risk management that is emphasized in the United Nations Sendai Framework for Disaster Risk reduction (UNDRR 2015) and the new EU directive (CER Directive 2022), this study aims to improve the risk and resilience assessment of power grid assets, by conducting a comprehensive literature review of fragility assessment models against critical climate hazards at a global level. In particular, the objectives pertain to: (i) the evaluation of causes, contributing factors, and consequences of previous failure incidents, with a view to providing information on the most critical hazards to enhance decision-making for investments in climate adaptation, (ii) the identification and classification of robust fragility models based on the failure mode, region, hazard, and asset type, enabling risk analysts to identify the most suitable models and reduce uncertainty in risk assessments with a focus on assets of the transmission and distribution network; (iii) the exploration of methods to incorporate climate change effects in fragility and risk assessment; (iv) the identification of potential applications of FCs in risk management of power grids. Recently, Schweikert and Deinert (2021) called upon research in power systems to clearly distinguish risk and resilience management concepts and to standardize the format of data and risk assessment methods for uncertainty reduction. This urging fully aligns with the aforementioned objectives.

## 1.2 | Review Methodology and Bibliometric Analysis

A systematic literature review was carried out by accessing the Dimensions database (Dimensions 2018). The database was selected, because it is robust (about 25% more publications compared to other databases), and provides a user-friendly interface. The review was conducted in four steps: identification, screening, eligibility, and selection, with the objective to identify publications that propose fragility models for power grid assets. Due to the scarcity of publications earlier than 2006, this review included records published between 2006 and 2023 (Figure 1a) to thoroughly analyze recent developments in the field. Initially, it was decided to cluster the keywords based on the terms "fragility," "power grid asset" and "natural hazard." Based on the most relevant identified publications and the authors' experience in the field, several keywords were inserted in each cluster. This resulted in the following full search string: (fragility OR "damage function" OR "damage model") AND ("power grid" OR "electricity grid" OR "electrical grid" OR "electric grid" OR "electric power" OR "power system" OR "overhead line" OR "transmission line" OR "transmission network" OR "transmission line" OR "transmission tower" OR "transmission system" OR pylon OR "distribution network" OR "distribution system" OR "pole" OR "distribution line" OR substation) AND ("natural hazard" OR climate OR drought OR earthquake OR "extreme weather" OR "extreme temperature" OR "high temperature" OR flood OR "sea-level rise" OR snow OR ice OR "multiple-hazard" OR "space weather" OR tsunami OR wildfire OR wind OR "cold wave" OR "heat wave" OR hurricane OR cyclone OR typhoon). The initial search of journal articles identified a total 361 records (identification step). After reading the abstracts, 173 items were selected (screening step). The subsequent analysis, which applied eligibility criteria by reading the full text of these screened records, returned 140 records (eligibility step). Publications were deemed ineligible for this study if they were preprints; conference proceedings; not accessible; or were not written in English. In addition, articles which: referred to geophysical hazards (e.g., earthquakes); assessed power grid resilience using fragility models developed by other researchers; or utilized empirical fragility models provided by electric power companies and HAZUS (FEMA 2013), were also excluded. This left 61 records deemed relevant for inclusion in the review. Therefore, the selected records included publications that specifically developed climate fragility models for power grid assets, ensuring the relevance of the findings to the objectives of the study.

Besides the systematic literature review, a bibliometric analysis was conducted to highlight gaps in the research field. To illustrate the increasing trend of resilience estimation, 140 eligible records were used in this analysis, including studies on geophysical hazards and resilience that do not propose new fragility models. To this effect, a co-occurrence network of the first 74 most frequent authors' keywords, appearing at least 5 times in 140 eligible records was visualized using VOSviewer software (VOSviewer 2023; Figure 1b). It was observed that the keywords were grouped into four clusters, and the most frequent keywords inside each cluster were "resilience" (1st cluster), "earthquake" (2nd cluster), "wind" (3rd cluster), and "pole" (4th cluster). The term "fragility" did not appear as a frequent keyword, because it often appeared as a phrase, such as "fragility curves" or "fragility analysis." The term "resilience" was identified 88 instances out of 140 records analyzed, surpassing the 77 instances of fragility.

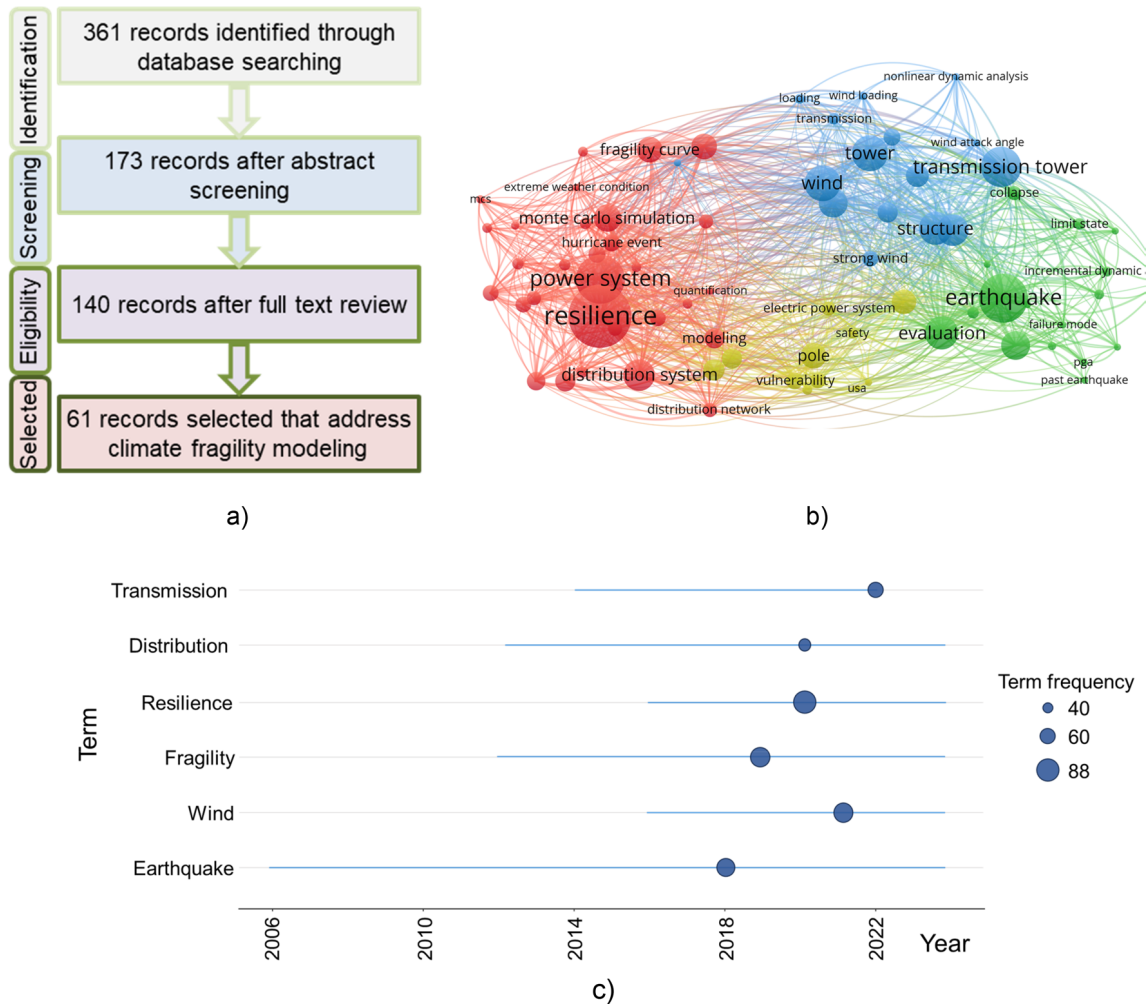

**FIGURE 1** | (a) The systematic literature review based on the open-access Dimensions database (Dimensions 2018); (b) keyword co-occurrence analysis of 140 eligible records; (c) unigram trending analysis of authors' keywords (the horizontal lines represent each keyword, and the position of the circle represents the average publication year).

A unigram trend analysis of the most frequent keywords clearly illustrated that resilience assessment studies outnumber those focusing solely on fragility. Furthermore, there has been an increasing trend in recent years toward assessing resilience instead of fragility, with an average publication year of 2020 compared to 2019 for fragility studies. This suggests that existing fragility models are increasingly being used for resilience assessment. However, it is crucial to determine whether these models are sufficient and fit for the purpose for addressing adaptation to climate extremes. Figure 1b,c shows that the majority of studies looked at fragility models on earthquake and wind hazards, while models for, for example, wildfires or floods (including climate change effects) were absent.

Interestingly, the equal distance of "transmission tower" from both "earthquake" and "wind" hazards in Figure 1b, suggests that this asset has been studied equally with respect to these two hazards, although wind has been found more damaging, according to the statistical analysis of Karagiannakis, Panteli, and Argyroudis (2023).

Another gap identified from the bibliometric analysis is the lack of consideration for multiple-hazard stressors in the fragility models. Scenarios involving combinations like wind and icing or flood followed by landslide or earthquake are absent in the records examined. These findings motivated the Authors to carry out a comprehensive review of fragility models in Section 4, and to highlight areas for further research.

## 2 | Typology of Power Grid Infrastructure

A power grid is a complex interconnected system that can be divided into three major sub-systems: (i) traditional (e.g., oil, coal, gas or hydroelectric) or non-traditional (e.g., wind farm or solar panel stations) power generation plants, (ii) transmission network, and (iii) distribution network (Cavalieri, Franchin, and Pinto 2014; Figure 2). The transmission and distribution networks incorporate electrical substations and loads depending on the voltage level. The purpose of a power grid is to deliver electricity from power plants to industrial and domestic customers. A transmission system is used to transmit electricity from a generation plant to multiple substations at very high voltage, which varies (in Europe) between 110 kV and 750 kV (ENTSO-E 2023a; Karagiannis, Chondrogiannis, et al. 2017). It comprises high-rise lattice towers—typically made of steel—which support overhead power lines or conductors. The typology of the towers and the distance between them depend on the environmental, topographical, and geomorphological conditions, as well as the voltage level, the number of circuits, and regional specifics. The towers are designed with two or more arms that support the lines, and insulators or bushings connect these arms with conductors. A ground wire is supported at the top for lightning protection.

Furthermore, substations are located inside generation plants and at numerous points throughout the grid. In the former case, substations use transformers to step up the voltage of electricity, enabling efficient long-distance transmission. At the end of a transmission line, there is another substation, which steps down the voltage and links the transmission line with a distribution network. Substations can be indoors or outdoors, and their layout and equipment vary significantly, depending on the type and region (Cavalieri, Franchin, and Pinto 2014). Once the voltage is stepped down, the electricity is transferred through the distribution network to customers at voltages that vary from tens of thousands down to 240 V in Europe (ENTSO-E 2023c). The power grid includes generation plants and customer loads, which are out of the scope of the present review. The interested reader can find more information on the different types of power generation plants and their resilience in an ad-hoc report by the World Bank (Nicolas et al. 2019), while further information on energy consumption at the European level can be found in relevant reports by EMBER-Climate organization (EMBER 2023; ENTSO-E 2023d). Finally, a review of underground cables was briefly conducted, since they are generally considered to be less vulnerable to natural hazards, and their application—as a mitigation measure—is limited to specific regions due to high cost.

## 3 | Climate Hazards and Previous Incidents of Power Grid Failure

### 3.1 | Major Power Grid Incidents of Failure

Table 1 shows representative major incidents of power grid failures associated with climate hazards, namely meteorological/climatological and hydrological. The purpose is to highlight the impact of these hazards on various power grid assets and the economy, including any damaging compound hazards as well as any successful preventive and recovery measures. The hazard events were selected based on the affected population, region, as well as recovery time. The first incident regards the 2005 Münsterland power blackout that was triggered by the compound hazards of ice and wind, which combined with aging effects increased infrastructure vulnerability (Klinger et al. 2011). Strong winds and heavy snowfall resulted in the accumulation of wet snow on overhead electrical lines in the form of snow rolls, leading to the failure of 82 high-voltage transmission towers. The power company that operated the transmission system in the region referred to the failure as a "black swan" event; however, news media (Spiegel 2005) and forensic analysis by Klinger et al. (2011) revealed that the collapse of the towers was caused due to the combination of excessive ice and wind loading along with the reduced capacity (structural strength) of the tower. The blackout affected more than 250,000 people for a period of 4–6 days, with estimated repair costs exceeding €130 million. This incident highlights the need for retrofitting or replacing aging infrastructure due to the exacerbation of climate hazards and for establishing a communication channel between power operators and meteorological agencies to identify compound hazards. Indeed, in October 2019, a power utility company in California shut off electricity to around 2.5 million

4 of 28

Wiley Interdisciplinary Reviews: Climate Change, 2024

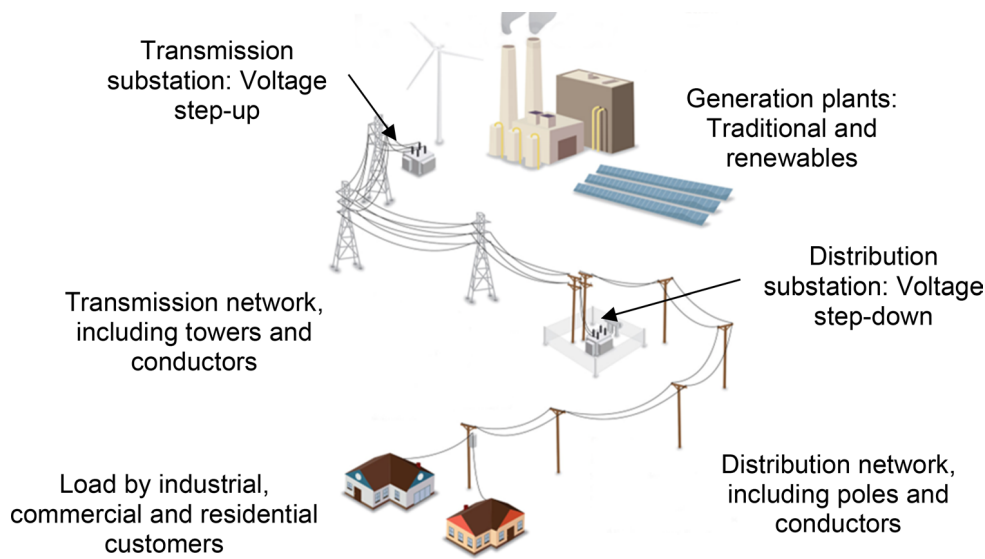

**FIGURE 2** | A typical electricity power grid, including power plants, substations, transmission, and distribution networks. Adapted from KSU (2021).

people to avoid deadly wildfires and damage to the power grid network, taking into account a weather forecast that predicted extreme weather conditions due to compounding heat and wind hazards (City-Journal 2019). Although a case-study carried out in the aftermath of the event demonstrated that the impact would have been catastrophic without the prior shutoff by the company, it also revealed that the preparedness of authorities and the resilience of the power grid were insufficient. It took over a month to restore the power in every region due to the high number of required inspections and high uncertainty regarding the condition of power grid assets (PG&E 2021). Furthermore, assets impacted by the heat wave could still trigger a wildfire when restoring power because of their proximity to vegetation.

Tropical cyclones, particularly hurricanes are among the most devastating hazards to the power grid. In 2017, Hurricane Maria severely affected the electricity system of Puerto Rico with the restoration of the power grid alone costing as much as €18 billion, the highest among all the incidents listed in Table 1. Due to the high vulnerability of the grid, only 20% of the transmission lines were restored after 1 month and the entire restoration process took over a year to complete (NYT 2021; Schweikert et al. 2019). Finally, the 2021 European floods that affected the German power grid demonstrated the severe effects of flood hazards on numerous pieces of equipment in substations for high impact-low probability events. The power outage lasted for up to eight-weeks due to site inaccessibility and repair work (News 2021; NYT 2021).

### 3.2 | Failure Mechanisms and Lessons-Learned

Table 2 provides a review of various causes of failure, contributing factors (e.g., environmental conditions), and failure modes for power grid infrastructure. The terms "failure cause" and "failure mode" are distinguished; the former addresses why an asset fails, while the latter explains how it fails. The purpose of this review is twofold: first to enhance the accuracy of FC modeling from an engineering perspective through a deeper understanding of key failure causes and failure modes, and second, to improve resilience management. For example, the wind attack angle significantly affects the structural response of transmission tower-lines, making it a critical factor to thoroughly examine in FC modeling. Also, increasing awareness and emergency planning, which are identified as contributing factors, can reduce vulnerability and enhance recovery capacity in case of an incident.

The review focuses on the energy security aspects related mostly to the dimensions of resilience and reliability, as classified in Sovacool (2011). Economic or socio-political dimensions are beyond the scope of this study, but more information on these aspects can be found in Cherp (2012). It is also noted that the list of factors in Table 2 is not exhaustive. Cherp (2012) illustrates that making a list of security concerns does not lead to a comprehensive understanding of the issue. This study could include additional causes and factors without a clear endpoint to this exploration. For example, the weight of falling trees on the lines should also be examined in the FC modeling. Nevertheless, contributing factors, such as tree density, age, and their position relative to power line and wind, are not included in the current study. In particular, the current list reflects the main and most representative causes of physical failure, and corresponding failure modes and contributing factors. These were identified in 328 incidents from two different major databases, as analyzed by Karagiannakis, Panteli, and Argyroudis (2023), as well as the systematic literature review described in Section 1.2. In both cases, machine learning has been trained based on the keywords and definitions of this study to identify these features.

Power grid assets present different levels of vulnerability to climate hazards. Based on previous failure incidents (see Table 1), multiple climate hazards can contribute to the failure of power grid assets. For instance, transmission towers are more susceptible to windstorms than floods, based on the statistical analysis of major incidents of failure by Karagiannakis, Panteli, and Argyroudis (2023). However, floods can cause scour of the foundation, which can amplify the effects of wind (Li et al. 2024;

5 of 28

**TABLE 1** | Some major incidents of power grid failure caused by various climate-related hazards worldwide.

| Hazard group | Main hazard(s) | Event, affected area, and reference(s) | Causes of power cut and contributing factors | Impact | Recovery time |
|---|---|---|---|---|---|
| Meteorological and climatological | Wind and heavy snow (compound hazards) | • 2005 Münsterland snowstorm<br>• Germany<br>• (Klinger et al. 2011; Spiegel 2005) | • Excessive ice and wind loading<br>• Aging effects | • 250,000 people without power<br>• 82,110 kV transmission lines ruptured<br>• Highway closure due to downed power lines<br>• €130 million total losses | Four to six days |
|  | Heat wave and wind (compound hazards) | • 2019 California wildfire<br>• USA<br>• (PG&E 2021; City-Journal 2019) | • Wildfire risk due to strong and dry winds caused pre-emptive shutoffs<br>• Increase of California's population in wildfire prone areas (sub- and ex-urban) | • Three million people affected and one death reported<br>• Water scarcity in pump-reliant areas<br>• Rail transport disruption for 2 days | More than 1 month |
|  | Wind (tropical cyclone) | • 2017 Hurricane Maria (Category 4)<br>• Puerto Rico<br>• (NYT 2021; Schweikert et al. 2019) | • Government underinvestment, poor vegetation management, and maintenance lapses in the electricity infrastructure<br>• Insufficient redundancy systems<br>• Only 15% of transmission lines designed for Category 4 hurricanes | • 70% of electricity customers without power<br>• $18 billion for power grid restoration (direct damage) | After 1-month, 20% of Puerto Rico's transmission lines were restored, while in some areas repairs extended to a year |
| Hydrological | Fluvial flood | • 2021 European floods<br>• Germany, Netherlands<br>• (News 2021; NYT 2021) | • Very high quantity of water—probably the highest in 1000 years<br>• Inaccessible substations due to damage to transport infrastructure | • 800,000 people without power at the beginning<br>• Four-days after the event, 102,000 people were still without power | Up to 8 weeks |

**TABLE 2** | Summary of failure causes, contributing factors, and failure modes for the main power grid infrastructure against climate hazards.

| Climate hazard | Asset affected | Common causes of power outages and representative contributing factors | Typical failure modes |
|---|---|---|---|
| Windstorm | • Transmission tower-lines<br>• Distribution pole-lines<br>• Outdoor substations | • High wind loading and wind attack angle (Scherb, Garrè, and Straub 2019)<br>• Flying debris and vegetation management, for example, falling trees, branches, tree types, cutting practices (Schweikert et al. 2019)<br>• Soil and vegetation, for example, trees with shallow rooting systems and soil with low drainage<br>• Lack of awareness and emergency planning, for example, communication protocols and training activities for capacity building<br>• Multi-hazard stressors (e.g., rainfall, wind, and catenary load due to ice) that can cause fatigue and galloping (Dong et al. 2022)<br>• Lack of monitoring and early warning systems<br>• Aging effects, for example, corrosion and fatigue (Liu and Yan 2022)<br>• Saltwater deposits in coastal areas (Karagiannis et al. 2019) | • Bending failure of towers (usually in the middle third of tower's height; Cai et al. 2019)<br>• Collapse of towers, heavy equipment, transmission and distribution lines due to debris, and fallen trees/branches<br>• Failure of the tower foundation or foundation-to-tower connection (Li et al. 2024)<br>• Damage to tower hardware, flashover and tripping<br>• Onset of different failure modes due to multiple hazards<br>• Electrical equipment failure due to the accumulated amount of airborne saltwater on equipment |
| Lightning strikes | • Transmission tower-lines<br>• Distribution pole-lines<br>• Outdoor substations | • Direct strikes and insulation flashover (Souto, Taylor, and Wilkinson 2022)<br>• Induced voltages from nearby strikes (Cao et al. 2022)<br>• Asset management, for example, aging effects, lack of periodical inspection, repair, and maintenance (Souto, Taylor, and Wilkinson 2022)<br>• Location with high keraunic levels<br>• Inadequate design criteria of lightning protection system, for example, insulation, lightning arresters, lightning conductors, power surge protection devices, and circuit breakers (IEEE 1410 2011)<br>• Lack of waterproofing measures<br>• Long transmission lines and large substations | • Short-circuit arcing fault (Souto, Taylor, and Wilkinson 2022)<br>• Shield wire or arrester failure<br>• Line tripping in a number of circuits |

(Continues)

| Climate hazard | Asset affected | Common causes of power outages and representative contributing factors | Typical failure modes |
|---|---|---|---|
| High temperature, heat waves, and wildfires | • Transmission tower-lines<br>• Distribution pole-lines<br>• Outdoor transformers or substations | • Peak loads increase due to high temperatures (Nazaripouya 2020)<br>• Reduction of transmission lines thermal capacity<br>• Excessive radiating heating from flame front (Serrano, Panteli, and Parisio 2023)<br>• Ionized air because of high temperature<br>• Soot and smoke<br>• No vegetation clearance and safety distances<br>• Loss of load-bearing capacity and ground instability from permafrost thawing, for example, subsidence, settlement, reduction in adfreeze strength, formation of talkiks and the thermokarst, thermal erosion, and mass wasting and increased corrosion (Tavares da Costa and Krausmann 2021)<br>• Inaccessibility of the location to fire fighters and complex topography | • Physical damage to tower, poles, conductors, and transformers (Nazaripouya 2020; Serrano, Panteli, and Parisio 2023)<br>• Foundation or structural failure due to soil settlement or subsidence<br>• Reduced capacity of lines and short circuits (Nazaripouya 2020)<br>• Deterioration of insulation properties<br>• Arcing between lines or between lines and the ground<br>• Transformer overheating, electrical paper degradation, and short circuits |
| Heavy rainfall and floods (river and flash) | • Electrical equipment inside substations<br>• Outdoor substations<br>• Transmission towers<br>• Distribution poles | • Exceedance of design flood level (National Grid 2016)<br>• No flood defense measures, for example, elevation, anchoring, strengthening, flood barriers and gates, drainage system, guide bunds, levees, diversion channels, or pumping stations (Krausmann et al. 2011)<br>• Lack of scour protection, for example, pile foundation, and corrosion resistance, for example, coating and inhibitors (Tavares da Costa and Krausmann 2021)<br>• Change in soil properties and susceptibility to landslides, for example, clay, silt, or sandy soils with low drainage, and no landslide protection measures, for example, slope stabilization with vegetation (McKenna et al. 2021)<br>• Vegetation type, for example, trees with shallow rooting system<br>• Poor maintenance of structures, equipment, and drainage systems | • Damage to equipment due to water drag forces, detachment of attached equipment due to buoyancy, and floating debris (Krausmann et al. 2011)<br>• Short circuit or electrical malfunction of equipment due to direct contact with water<br>• Transmission tower collapse or tilt due to settlement, landslide, or foundation scour |

Wahlin, Davis, and Kandaris 2011). Also, the power grid assets are ranked by their vulnerability against each hazard in the second column of Table 2. However, it should be clarified that a comparison of vulnerability may not always be possible due to under-reporting (Karagiannakis, Panteli, and Argyroudis 2023) and the type of reporting criteria. For example, in the case of lightning, transmission towers may not always exhibit higher overall vulnerability, because incidents of lower-impact failures may not be consistently recorded in the databases. Therefore, transmission lines are placed higher in the rank due to their higher impact in case of failure. Souto, Taylor, and Wilkinson (2022) demonstrated that transmission towers are more vulnerable to lightning than substations. The vulnerability among power grid assets is more straightforward in cases of floods and heat waves or wildfires.

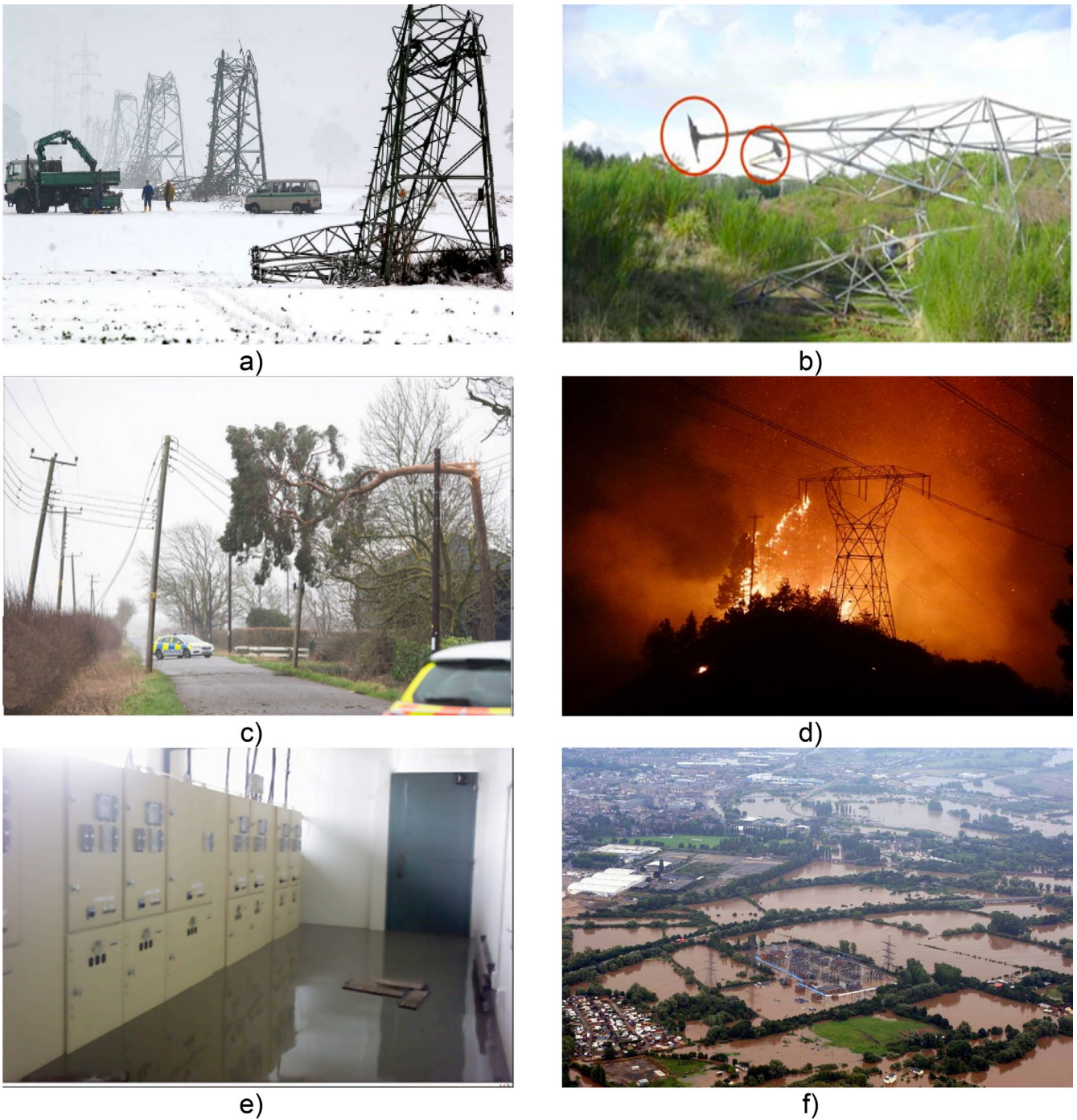

**FIGURE 3** | Failure modes of power grid sub-systems against natural hazards: (a) tower collapse due to compounding ice and wind (Spiegel 2005); (b) footing failure of the tower due to wind (Eidinger and Kempner 2012); (c) damage to distribution network due to fallen trees (BBC 2020); (d) damage to transmission network due to forest fire (RTO 2021); (e and f) equipment malfunction due to indoor and outdoor substation inundation (Eidinger 2011) and (National Grid 2016).

Inadequate design criteria, exceedance of design levels, insufficient maintenance, as well as lack of risk management, awareness, and emergency planning are common failure causes and contributing factors among all hazards, as demonstrated in major incidents of Table 1. Windstorms affect mostly the transmission and distribution networks due to the high-induced wind loading. In addition, compounding-induced hazards such as wind with ice or rainfall can contribute to tower/pole failure and galloping phenomena (Dong et al. 2022). In such cases, failure due to bending or member buckling can occur in the middle third (Figure 3a) or bottom of a tower (Figure 3b; Cai et al. 2019; Liu and Yan 2022; Zhang, Song, and Shafieezadeh 2022).

Buckling and yielding in cross arms and leg members in the middle tower section are predominantly observed in multiple hazard scenarios, for example, when an earthquake follows a typhoon (Jeddi et al. 2022). It should be noted that wind angle is a critical parameter that affects substantially the wind impact. Karagiannakis, Panteli, and Argyroudis (2023) showed that the most common cause of failure of structures and equipment relates to falling trees and airborne debris (Figure 3c), due to poor vegetation management, unsuitable trees for a wind-prone region with shallow rooting system, and soil with low drainage. Contributing factors such as wind fatigue and corrosion on towers have also been observed. Furthermore, equipment in outdoor substations can fail due to the corrosive effects of airborne water spray in coastal areas (Karagiannis et al. 2019).

Lightning flashes can directly or indirectly impact the power grid and result in short-circuits (Souto, Taylor, and Wilkinson 2022). When the strike is direct, it causes insulation flashover due to the excessive voltage that enters the system, surpassing insulation levels. Long-distance transmission lines and large substations located in areas with high lighting activity or at high altitudes where taller objects such as trees and buildings are scarce, increase the exposure and impact on the power grid. Additionally, induced voltages from lightning strikes that hit the ground in proximity to power lines are also a common cause of outages (Cao et al. 2022). On the other hand, inadequate design criteria of lightning arresters and conductors or power surge protection devices and circuit breakers are less frequent causes of failure (IEEE 1410 2011).

Power outages caused by high temperatures, heat waves, and wildfires stem from both structural and non-structural damage to power grid assets. First, high temperatures lead to a significant increase in electricity consumption, resulting in higher peak loads in terms of magnitude and duration. This, in turn, causes transformers to overheat, reduces the capacity of lines, and eventually, leads to short circuits due to line sagging (Nazaripouya 2020). On the other hand, wildfires can physically damage electricity grid assets through the flame front, smoke, and ionized air (RTO 2021; Serrano, Panteli, and Parisio 2023). This can result in the collapse of wooden poles, particularly in areas with poor vegetation management (Figure 3d). Lastly, high temperatures resulting from climate change can lead to permafrost thawing in Artic regions. The thawing can decrease the load-bearing capacity of the soil, causing foundation failure due to soil settlement or subsidence (Tavares da Costa and Krausmann 2021).

When it comes to floods, the primary impact is on the equipment of indoor or outdoor substations due to the exceedance of design inundation levels or lack of flood defense measures, for example, flood barriers and gates, guide bunds, levees, drainage systems, and pumping stations (National Grid 2016; Figure 3e,f). For example, floating debris or the buoyancy of electrical equipment can cause collisions, leading to the collapse of towers and other equipment within substations (Krausmann et al. 2011). Also, poor maintenance for scour protection (e.g., pile foundation) and corrosion resistance (e.g., coating or inhibitors) can lead to foundation failure and the collapse of transmission towers. Furthermore, changes in soil properties due to moisture ingress (McKenna et al. 2021)—which reduces soil strength and increases susceptibility to landslides depending on soil type (e.g., clay, silt, or sandy soils)—and the lack of protection measures for slope stabilization (e.g., vegetation), can result in tower or pole failure. In particular, scour-induced failure and unsuitable vegetation were the contributing factors to the failure of towers and pylons during the 2023 floods in Emilia-Romagna, Italy (ReFLOAT-ER 2023) and Thessaly, Greece (CNN 2023).

## 4 | Fragility Analysis Models

### 4.1 | Importance and Definition

Fragility curves (FCs) are practical tools that can be used for the probabilistic risk assessment and management of critical infrastructure—such as power grid and transportation assets or process plants—exposed to climatic and other hazards. FCs are also employed in conducting stress tests and impact analyses relating extreme events to critical infrastructure (Argyroudis, Fotopoulou, et al. 2020; Lam et al. 2018). They were initially used in the seismic risk assessment of nuclear power plants (Kennedy and Ravindra 1984), while their use has been expanded to include other infrastructures, for example, transport (Argyroudis, Mitoulis, et al. 2020; Palin et al. 2021), power grids (Serrano-Fontova et al. 2023) or process plants (Karagiannakis et al. 2022; Di Sarno and Karagiannakis 2020) as well as other hazards, for example, climatic (Panteli and Mancarella 2017; Nazemi et al. 2023).

The graphical representation of an FC depicts the probability that a structural system exceeds a certain level of physical damage (e.g., minor, moderate, extensive, complete) given an Intensity Measure (IM) of a natural hazard such as wind speed, water depth, or fire height. The level of damage is typically classified based on Damage States (DSs) using an Engineering Demand Parameter (EDP). The EDP corresponds to the induced response or required strength due to external loads from natural hazards. Examples of EDPs for power grid assets include the tower tip displacement, the bending moment of a truss element, or global damage accumulation of a transmission tower. In some cases, the "damage" is expressed as the performance of the power grid, measured by performance indicators such as the percentage of the total power network length affected or the number of customers without electricity. Furthermore, the impact of climate change on the power grid can be reflected in the FCs. If an asset is designed for an IM of a climate hazard, $IM_1$, and this IM increases at $IM_2$ due to climate change, the probability of exceedance of a certain DS for the as-built power grid assets increases from $P_1$ to $P_2$ (Figure 4a). Another method pertains to the shift of the FC to the left due to damage accumulation—for example, foundation scour—which is aggravated by climate change. This change in the fragility can be assessed using numerical models accounting for aging effects (see examples in Figure 7a). In this case, the probability of exceedance will also be higher $(P_2 > P_1)$ even for the same IM, $IM_1$ (Figure 4b).

Traditionally, the lognormal distribution has been used extensively to describe efficiently the EDP-IM relationship of single hazard scenarios, for example, wind (Alipour and

10 of 28

*Wiley Interdisciplinary Reviews: Climate Change*, 2024

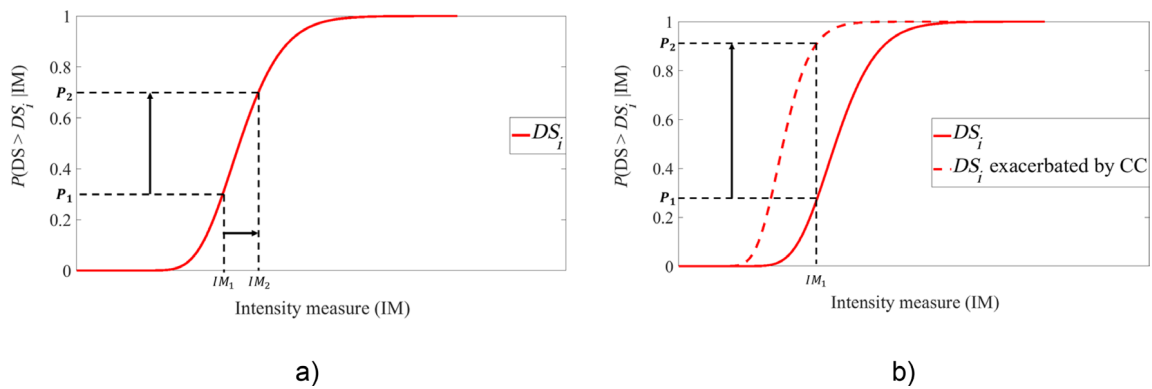

**FIGURE 4** | Two methods to reflect the impact of climate change (CC) on the fragility curves for a specific damage state, $DS_i$: (a) increase in intensity measure (IM); (b) shift of the fragility curve resulting from damage accumulation.

Dikshit 2023; Bjarnadottir, Li, and Stewart 2013; Reed and Wang 2018) or flood (Argyroudis and Mitoulis 2021; Prasad and Banerjee 2013). Most of the produced FCs are one-dimensional, which means that a single IM is used to predict the probability of structural damage. However, adopting a multi-hazard parameter approach results in a more accurate representation of structural damage (Cimellaro et al. 2006). For example, Koutsourelakis (2010) considered single and compound seismic IMs to examine the response of a geotechnical problem and inferred that the combination of two IMs provides a superior prediction. By virtue of limitations of the lognormal distribution to capture efficiently the data that pertain to multiple-parameter events or multiple-hazard events, the logit function was adopted instead. This statement is also supported by Jeddi et al. (2022) and Reed, Friedland, et al. (2016) who employed logit functions for power grid assets against multiple hazards, for example, wind and rainfall or rainfall and inundations. In fact, the selection of the function for the fragility description is another source of uncertainty, the consequence of which can be determined in comparison with empirical data (Bakalis and Vamvatsikos 2018).

The methods used for deriving FCs are categorized mainly into empirical, judgmental, analytical, and hybrid (Schultz et al. 2010), depending on the way that data are collected. Empirical FCs are formed using observational data that are systematically monitored, controlled, and stored. Judgemental data refer to the elicitation of expert opinions through surveys, to supplement insufficient observational data. The data may include various modeling parameters correlating the level of hazard intensity with the expected damage levels; however, the quality of collected data heavily relies on the experts' experience, making bias more challenging to control. Thus, the empirical method is considered more realistic, as it can incorporate various structural response factors observed during post-extreme weather surveys. However, empirical FCs tend to be scarce due to the limited availability of data, primarily within the low hazard intensity range.

In contrast, analytical methods rely on physical models or explicit demand–capacity relationships. They can encompass different structural configurations, geotechnical characteristics, or environmental factors. While analytical FCs can reduce biases referring to material and hazard uncertainty, include all possible failure modes, and yield robust risk assessments, careful consideration is necessary during modeling due to limitations in software modeling capabilities. Given the demanding and onerous nature of the modeling process, the validity of analytical FCs can be verified by comparing them with other relevant FCs. However, this is not always possible due to the dissimilarities in structural layout, soil properties, and climatic conditions. Consequently, hybrid fragility investigation can be conducted to address modeling challenges, combine diverse data sources, and minimize both modeling and hazard-induced uncertainties (Silva et al. 2019).

Furthermore, analytical FCs are decomposed into different solution methods—analytical and numerical. Numerical methods have gained prominence in recent years over analytical ones to overcome simplifying assumptions of the analytical approaches and the reduction of computational cost. The finite element method, for example, is generally considered a computationally demanding method for complex structural systems. The most robust numerical solution method is Monte Carlo simulation (Cai et al. 2019; Ma, Bocchini, and Christou 2020; Ma, Dai, and Pang 2020; Scherb, Garrè, and Straub 2019; Xue et al. 2020), whereas the Latin Hypercube Sampling (Cai and Wan 2021; Liu and Yan 2022; Pan, Li, and Tian 2021) is considered a reliable alternative for saving computational time, for example, for sampling structural and material properties. More information regarding the benefits and shortcomings of each solution method can be found in Schultz et al. (2010). Additionally, machine learning methods have emerged as powerful numerical approaches, capable of analyzing complex systems by considering large datasets (e.g., monitored, numerically derived) to predict outcomes and optimize decision-making (Kazantzi et al. 2024). These methods can complement (Watson et al. 2024) or even replace (Hughes et al. 2022) analytical methods, providing enhanced accuracy and efficiency in risk assessments.

The following sections, present a comprehensive review of climate fragility models for power grid assets aiming to highlight strengths, identify limitations, and explore potential directions for future research. The available models are categorized based on different assets of the power grid and further classified according to the associated climate hazards. All the fragility models found in this review that were developed within the last two decades are summarized in Table A1 of Appendix A, including asset type, associated natural hazard, region of the study, fragility model (e.g., analytical, empirical), intensity measure, reference and year of publication.

## 4.2 | Transmission Towers and Lines

From Table A1 of Appendix A, it is evident that transmission towers coupled with transmission conductors have received more attention in fragility studies compared to other subsystems of the power grid. The focus has primarily been on wind loading, because transmission towers are high-rise, relatively lightweight and brittle structures, and thus more vulnerable to meteorological phenomena than geophysical phenomena. Also, they are more influential, since they transmit electricity to a larger population compared to distribution poles. Notably, based on the exclusion criteria defined in Section 1.2, only one study (for Europe) has been found that derives FCs for wind hazard. Also, one study has been identified to address multiple hazards in Europe and the US. It should be noted, though, that there are studies (e.g., Klinger et al. 2011; Liang et al. 2015; Shehata, El Damatty, and Savory 2005; Tibolt et al. 2021), among others, that focus on designing and assessing transmission towers and their failure modes according to relevant structural design codes (EN 2005).

### 4.2.1 | Wind

Previous incidents revealed that transmission towers are more fragile than conductors as the latter are designed with high safety margins (Ma, Bocchini, and Christou 2020; Ma, Dai, and Pang 2020). This was also confirmed by the analytical study of Rezaei et al. (2017), albeit several other researchers have reported higher fragility of conductors compared to towers based on empirical data and expert judgment (Alipour and Dikshit 2023; Panteli et al. 2017; Panteli and Mancarella 2017), as shown in Figure 5a. Empirical data may account for the failure of lines due to fallen trees, which justifies the higher fragility. Brown (2009) examined the failure rate of transmission towers due to hurricanes over a 10-year period within a 50 km radius of the Texas coast. The data was adequate to examine the performance of towers across different categories of hurricanes. However, the availability of such extensive data is not as prevalent in other regions, for example, Puerto Rico (Dos Reis et al. 2022) or Louisiana state (Reed, Powell, and Westerman 2010). This scarcity is attributed to under-reporting, stemming from the unique challenges associated with documenting natural hazard events affecting local regions. This under-reporting can introduce bias and render studies more region-specific.

The limitations of empirical studies can be addressed by analytical fragility models. As illustrated in Figure 5a, analytical FCs present less dispersion compared to empirical FCs, since they account for specific structural typologies and region-specific climatic variables, for example, wind speed and angle (Alipour and Dikshit 2023; Chen et al. 2022; Scherb, Garrè, and Straub 2019). Furthermore, analytical studies (Fu et al. 2019; Cai et al. 2019) confirmed also by others (Liu and Yan 2022; Ma, Khazaali, and Bocchini 2021; Rezaei et al. 2017; Fu and Li 2018) highlighted the significance impact of the attack angle and span length on tower performance. Figure 5b illustrates that smaller span lengths and wind direction perpendicular to the line result in higher fragility. Apart from the span length, Hou and Chen (2020) evaluated the probability of failure of a transmission line considering the probability of trees' failure along the line. While several researchers, for example, Liu and Yan (2022) and Fu et al. (2019), adopted tip displacement as an engineering demand parameter, Ma, Khazaali, and Bocchini (2021) corroborated that an asset-based fragility model is more robust, since it accounts for the progressive global collapse of towers through element removal. However, the study considered only the collapse damage state, which is insufficient to characterize the structural state and prioritize inspection in the aftermath of a severe natural hazard (Fu et al. 2019). Most studies (Cai et al. 2019; Liu and Yan 2022; Panteli et al. 2017; Panteli and Mancarella 2017; Rezaei et al. 2017), among others, have focused only on the collapse damage state to correlate tower collapse with electricity outage.

Analytical FCs can also effectively capture both epistemic and aleatoric uncertainty, which derives from the variability in capacity (e.g., uncertainty in structural properties such as yield strength or angle steel thickness) and demand (e.g., uncertainty in environmental actions including climate change effects), respectively. While fragility analysis commonly considers only aleatoric uncertainty, accounting for uncertainty in structural parameters widens the dispersion of fragility (Fu et al. 2019). Other studies have estimated that the wind resistance of transmission towers can decrease by up to 20% over a 25-year period when exposed to environmental conditions that accelerate corrosion (Liu and Yan 2022).

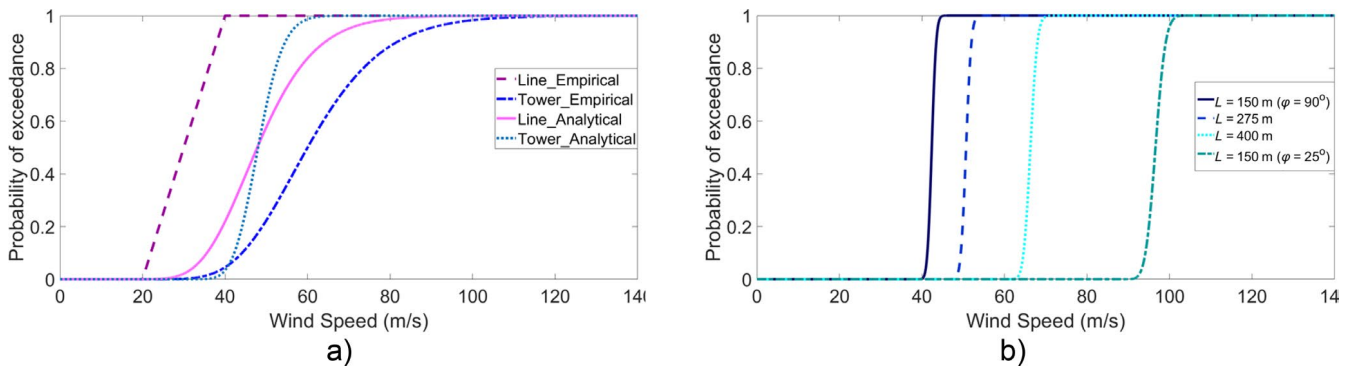

**FIGURE 5** | Comparison of fragility models for transmission towers and lines against wind hazard: (a) empirical and analytical FCs (Dos Reis et al. 2022; Alipour and Dikshit 2023; Scherb, Garrè, and Straub 2019); (b) FCs for different span length ($L$) and wind attack angle ($\varphi$; Cai et al. 2019).

## 4.2.2 | Heavy Snow and Lightning

There are few available fragility models for transmission tower—line systems with respect to the single hazards of heavy snow and lightning, which mainly assess the probability of failure based on empirical and meteorological data. Lu et al. (2018) proposed a resilience assessment methodology for ice disasters affecting transmission systems by considering a cell partition method for spatial resolution of meteorological data and an empirical fragility model of ice accretion. To determine the probability of transmission system failure, a Monte Carlo simulation was employed, assuming an exponential FC and setting a threshold value of 15 mm for the ice thickness. The main advantages of the proposed model are the high resolution of storm paths achieved through the partition method and the consideration of impact duration, which increases the accuracy of the model.

In line with Lu et al. (2018), Solheim and Trotscher (2018) highlighted the importance of using finely meshed meteorological data for the impact of lightning on transmission systems. In particular, to estimate the hourly probability of failure, two steps were followed; first, a long-term failure rate was calculated using a Bayesian approach and historical data, and second, the hourly probability of failure was evaluated by considering the lightning exposure at any given time. The exposure was estimated based on two lightning indices, namely K and TT, which depend on temperature, dew point temperature, and surface pressure. In contrast to Solheim and Trotscher (2018), Bao et al. (2021) expressed the fragility of a transmission system trip-out as a function of thunderstorm duration, enabling power companies to estimate the probability of lightning-related trip-out and implement pre-emptive measures for risk mitigation. Arguably, to understand the impact of lightning on power grid assets requires further research, as it remains one of the most damaging hazards.

## 4.2.3 | Multiple-Hazards

Transmission towers can be exposed to multiple dependent or independent hazards. Dependent hazards refer to meteorological hazards that appear simultaneously, for example, wind accompanied by rain or snow, whereas independent hazards refer either to meteorological, climatological, or geophysical hazards that are not accompanied or affected by the appearance of another hazard in time, for example, earthquake followed by strong winds (Leonard et al. 2014). Dependent hazards are modeled as concurrent events in structural analysis, whereas independent hazards can be concurrent or non-concurrent. In contrast with conventional FCs, multiple hazards are described by fragility surfaces that account for the coupling between two IMs. It should be noted that existing studies use either structural analyses to derive multi-surface FCs or empirical models based on power outage data provided by electricity distribution operators. Although multiple-hazard scenarios are of great significance in windstorm- and hurricane-prone regions, exacerbated by climate change, the literature contains only a limited number of studies.

In terms of dependent hazards, Fu, Li, and Li (2016); Fu et al. (2020) and Bi et al. (2023) examined the response of a transmission tower subjected to the simultaneous impinging forces of wind and rain load. Fu, Li, and Li (2016) estimated that the probability of collapse damage state exceedance was very close to zero for the design wind speed (42 m/s) which corresponded to a rain rate of 619 mm/h. Undoubtedly, rain rates higher than 619 mm/h are rarely observed, thus rainfall was less critical for this study. In fact, different tower typologies exhibit varying fragility levels; for example, Fu et al. (2020) inferred that the collapse wind speed for a rain rate of 240 mm/h was 24 m/s, whereas Fu, Li, and Li (2016) found a considerably higher wind speed (greater than 42 m/s) for the same rain rate. In contrast with the previous studies, Snaiki and Parida (2023) developed numerous synthetic hurricane records for both present (1990–2020) and future RCP 8.5 (2020–2050) climate scenarios. The future database was generated using CMIP6 projections of tropical cyclones from the high-resolution CMCC-CM2-VHR4 model. For this purpose, the transmission-line system of different cities along the US Atlantic coast was simulated using a finite element model, and a significant rise of up to 200% in the probability of failure was observed due to the compound wind and rain hazards. The increase varied considerably due to the substantial change in the hurricane intensity and frequency along the US east coast as a result of the changing climate effects.

Additionally, Rezaei et al. (2017) focused on ice accretion and wind speed and evaluated the combination of these hazards along with wind angle that resulted in the failure of a transmission network with the highest probability. In Figure 6a, the FCs at the wind speed with the highest likelihood (design speed) and at different values of ice thickness are presented. As can be seen, both the tower and lines, are considerably more fragile when $\varphi = 90°$ (angle relative to the direction of the lines) due to the ice accumulation on the lines.

When it comes to independent hazards, only wind and earthquake have been examined in literature in relation to transmission towers. These hazards can occur simultaneously or within a time period in which repairs from the first hazard are not yet completed. Li, Pan, et al. (2022) investigated the multi-hazard (wind and earthquake) performance of a transmission tower over its lifetime by considering long-term wind-induced fatigue, which is a major concern for high-rise steel transmission towers. To account for wind-induced fatigue, a continuous degradation of mechanical properties over time was considered, and highlighted that the leg members at tower bottom are more susceptible, as they experience the main overturning moments caused by wind. This outcome was also confirmed by Jeddi et al. (2022) who examined a transmission tower subjected to earthquake and typhoon impacts. Also, Li, Pan, et al. (2022) developed a critical safety surface that corresponds to 10% probability of exceeding the collapse damage state and relates the peak ground acceleration (PGA) with the basic wind speed (i.e., 3-s gust wind speed with a specific return period) for different time points of fatigue level (in years). The critical surface was produced as an effective and straightforward tool for fragility evaluation by risk analysts. In contrast to the previous study that accounted for the concurrent occurrence of wind and earthquake hazard, Jeddi et al. (2022) analyzed a transmission tower, while taking residual damage from the previous hazard into account. The study generated numerical models using the Latin Hypercube Sampling method, while machine-learning techniques, specifically gradient boosting, were used for the classification and the logistic regression of the collapse dataset. Figure 6b shows the fragility surface of the earthquake-typhoon multi-hazard scenario, which was significantly more damaging than the typhoon-earthquake scenario. Nevertheless, the effect of the prior hazard in both scenarios was considerable, underscoring the need to consider multiple hazards in risk assessments.

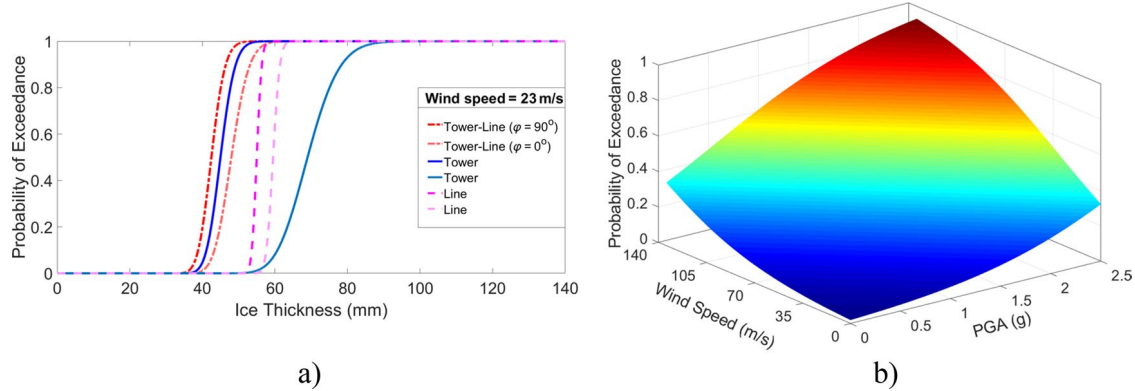

**FIGURE 6** | Fragility models for transmission towers and lines accounting for multiple-hazards: (a) concurrent ice hazard and wind hazard at design wind speed (Rezaei et al. 2017) and (b) concurrent wind and earthquake hazards (Jeddi et al. 2022). PGA, peak ground acceleration.

## 4.3 | Distribution Poles and Lines

Similar to transmission towers, the majority of available fragility models for distribution poles and lines focus on wind hazard, as the most common causes of failure including falling trees and airborne debris (Najafi Tari, Sepasian, and Tourandaz Kenari 2021; Shafieezadeh et al. 2014). Poles are light columns that primarily support the weight of the lines and are less vulnerable to earthquakes, which explains the low number of fragility studies conducted for this hazard. Regarding multiple hazards, most of the available studies have been conducted in North America; this is likely because the majority of the distribution poles are aboveground, as opposed to Europe. Yet, the fragility analysis of this asset has not received extensive attention in the literature.

### 4.3.1 | Wind

When wind-induced damage data is accessible for a distribution network, it is typically provided by the network operators and refers to power faults, the number of people affected, and restoration time. For example, in a study by Dunn et al. (2018), 12,000 wind-related power faults in the UK distribution network were analyzed to develop fragility models. The high spatial resolution improved the accuracy of fragility estimation. Improving the resolution of climate-related hazards can also enhance the accuracy of evaluating aging effects, which are affected by factors such as wind and humidity conditions. Also, Najafi Tari, Sepasian, and Tourandaz Kenari (2021) adopted different FCs for medium voltage poles and conductors, giving emphasis to the application of region-specific coefficients, which vary with the geographical condition, age, type, and class of the pole. Dunn et al. (2018) stated that some slight modifications to these coefficients may be needed to apply FCs to other regions, especially at wind speeds up to 37 m/s, given that the majority of faults pertained mostly to the failure of lines. Additionally, Zhang et al. (2023) and Lu and Zhang (2022) produced fragility surfaces as a function of wind speed, pole height, and wind angle.

Besides pole and conductor typologies, the structural capacity of poles can vary across regions due to aging effects. Therefore, numerous analytical studies have focused on aging effects for different pole classes (ANSI 2015). For instance, Darestani and Shafieezadeh (2019), Dehghani, Mohammadi, and Karimi (2023), Lee and Ham (2021), Onyewuchi et al. (2015), Salman and Li (2016) and Shafieezadeh et al. (2014) derived age-dependent FCs based on the wind-induced moment demand and capacity, while they emphasized that fragility derivation should account both for aging and structural parameters such as structural typology (height, number of conductors, span length) and weather conditions (wind speed and angle). Figure 7a demonstrates FCs for two common classes (C3 and C5) and different time periods, and confirms the aforementioned statement. Parameterized FCs, which can facilitate risk and resilience assessments for a portfolio of assets, were developed by Darestani and Shafieezadeh (2019) by fitting a response surface to different sets of design variables (e.g., height or span length) and wind speed, providing a time efficient and robust tool for network-level resilience assessment. Additionally, Hughes et al. (2022) and Watson et al. (2024) demonstrated that machine learning can enhance the accuracy of physics-based FCs by considering historical data for model training.

Furthermore, due to the simpler layout of distribution poles compared to transmission towers, most analytical studies have relied on a bending moment–based failure criterion, which represents the maximum value within the ground and near the surface. Both Zhang et al. (2023) and Shafieezadeh et al. (2014) studied typical classes of wood poles in the US and considered an empirical relation for the moment capacity. Also, the wind-induced demand was calculated based on guidelines provided by ASCE (2013), accounting for uncertainty in initial strength, time-dependent capacity, geometric features of the wood poles, and lateral wind loads.

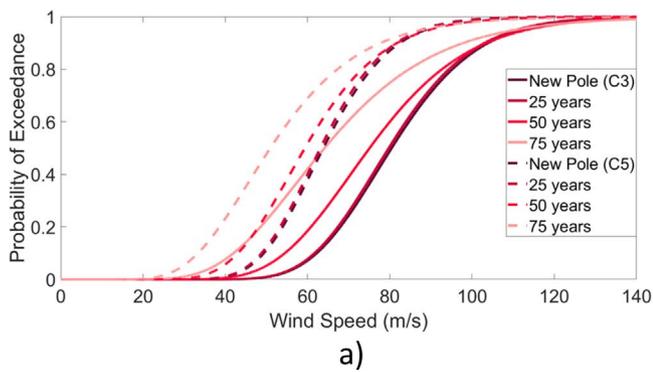 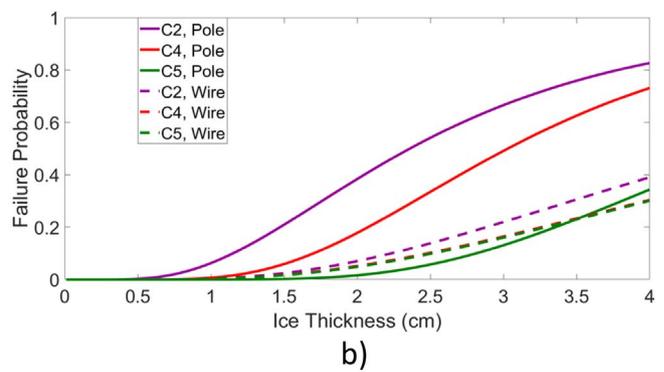

**FIGURE 7** | FCs for distribution poles, wires, and different pole classes (C2–C5) considering: (a) aging effects and wind hazard (Shafieezadeh et al. 2014) and (b) ice-hazard at 30 m/s wind speed (Hou et al. 2023).

### 4.3.2 | Multiple-Hazards

When it comes to multiple hazards, Reed, Friedland, et al. (2016) fitted two and three hazard data (wind, rainfall, and inundation) from hurricane Isaac in Louisiana, US, and concluded that the models incorporating interaction among the three hazards demonstrated the lowest dispersion based on the fitting criterion, which clearly illustrates the efficiency of the multi-hazard analysis concept. Ma, Bocchini, and Christou (2020) and Ma, Dai, and Pang (2020) assessed the concurrent impact of ice accretion and wind on the fragility of pole-wire elements at the component and system level. At the component level (pole or wire), low concurrent wind affected mainly the failure probability of the wire, whereas severe concurrent wind impacted both the wire and pole. At the system level, considering all pole-wire elements of the distribution network, the annual failure probability rose to 16%, which is almost three times greater compared to the failure probability of each component of the system. Hou et al. (2023) developed fragility surfaces for four failure modes, including failure of conductors and short circuit due to falling trees, but only for 1-phase lines (1 conductor and 1 neutral wire), as a falling tree would definitely cause a short-circuit for a 3-phase line. Figure 7b depicts the rise in the fragility of poles and wires at 30 m/s wind speed, as the ice thickness on wires grows up to almost 4 cm and the class of poles decreases from C5 to C2.

## 4.4 | Substations

The statistical analysis carried out by Karagiannakis, Panteli, and Argyroudis (2023) showed that flood hazard caused between 23% and 37% of incidents in power grid in substations in the examined disaster databases. Despite this high vulnerability, there are few models that evaluate the vulnerability of substations to floods and other climate-related hazards, for example, hurricanes, wildfire, extreme temperature, or sea-level rise.

### 4.4.1 | Flood

Methodologies for the evaluation of flood damage rely on empirical data or expert judgment. Damage functions are commonly used in flood risk analyses, assuming that once a substation is inundated, water causes 100% damage to all equipment at a certain inundation level. HAZUS (FEMA 2013) classifies substations into low, medium, and high voltage classes, assigning an overall rating of high vulnerability to all of them. It also proposes empirical fragility relationships derived from observations, correlating inundation levels with equipment functionality. A threshold of almost 1.2 m, representing 7% equipment damage, is applied to all classes. The damage percentage increases to 15% at the maximum considered flooding depth of approximately 3 m (the depth is measured from ground level, assuming that equipment is located about 0.91 m above ground). It is important to note that HAZUS considers damage to transmission or distribution lines only at elevated crossings (i.e., lines at ground level or below are considered as non-vulnerable), estimating this damage at 2% at roughly 3 m above ground level. However, there are limitations in using the HAZUS damage functions and applying them to regions other than the US due to the variation in substation standards, equipment, protective measures, and local conditions. This introduces additional epistemic uncertainty in flood risk assessments (Karagiannis, Turksezer, et al. 2017; Sánchez-Muñoz et al. 2020). To address these limitations, Karagiannis, Turksezer, et al. (2017) extended the depth-damage functions using polynomial and power functions, observing increased repair costs for substations when flood water depths exceeded 3 m in certain areas. Recently, Li et al. (2024) examined the fragility of pole-mounted substations against flood hazard. The study considered the flood depth, the water velocity, the scour of pole foundation, the concurrent wind speed, and floating debris as IMs. It was corroborated that debris contributes more to structural failure than scour. The most important limitations of the study referred to the constant angle of wind direction, debris attack angle, and the constant value of the scour for each flow velocity value.

## 5 | Risk Mitigation and Climate Change Adaptation Strategies

### 5.1 | A Risk Management Framework for Power Grid Assets

As stated in Schweikert and Deinert (2021), the concepts of resilience, fragility, and risk management have distinctive differences, but they are all important to maintain functionality in power systems despite disturbances. Presently, there is a growing interest in incorporating FCs as an indispensable element of methodologies for building resilience (Panteli et al. 2017;

15 of 28

Panteli and Mancarella 2015; Argyroudis 2022). The interest is justified by the need to employ practical vulnerability and resilience assessment models. Based on these models, different actors can quantify resilience and assess their preparedness, to account for challenges such as multiple-hazard scenarios (Li, Ahuja, and Padgett 2012; Turner et al. 2019), inter-dependencies among assets (Argyroudis et al. 2019), social and ecological concerns (Markolf et al. 2018) as well as climate change (Feng, Ouyang, and Lin 2022; Van Vliet et al. 2012). Box 1 and Figure 8 summarize the main applications of FCs within a resilience management framework of power grid assets, and highlight the multi-purpose nature of FCs, while also underlining the need for improved reliability of fragility modeling. To make the framework complete, processes and inter-sectoral actors involved in the decision-making are also stated. In the framework, all actors are linked and communicate with one another, which is considered necessary to achieve community climate resilience and sustainability. Indeed, the framework can be used as an adaptation measure itself, since it encourages the communication of inter-sectoral actors involved in the decision-making of power grid assets. More information on how the European Commission (EC), the US Department of Energy (DOE), or the UNDRR intend to strengthen the dialogue among the actors can be found in EC (2021), DOE (2016) and UNDRR (2023), respectively.

Special attention to climate change effects needs to be given to all actors involved in the framework. For example, risk engineers need to consider the change in the level of risk or probability of failure of power grid assets due to climate change and high-impact low probability events, as part of stress testing exercises (DeMenno, Broderick, and Jeffers 2022; Linkov et al. 2022). This can be achieved either directly through FCs or indirectly by estimating the increase/decrease in the annual probability of occurrence of a hazard scenario. For example, the first case was examined by Snaiki and Parida (2023), developing FCs based on a finite element model and synthetic hurricane databases that corresponded to RCP8.5. On the other hand, (Karagiannis, Turksezer, et al. 2017) considered the same flood hazard maps with several climate change scenarios from EURO-CORDEX models to evaluate the impact of the 100-year flood (reference scenario) in three-time slices, that is, 2020, 2050 and 2080 (EURO-CORDEX is a set of high-resolution regional climate models to downscale outputs from global climate models, providing detailed climate data for various future scenarios). The expected increase/decrease in the occurrence probability of the reference scenario was used to estimate the impact of climate change on the expected annual losses of a transmission grid network.

Power operators such as those involved in ENTSO-E in the EU need to ensure compliance with regulations. The EU Agency for the Cooperation of Energy Regulators (ACER) requires all transmission operators to establish a methodology for risk management (ACER 2019). This methodology will assist operators to go beyond the traditional "N-1 criterion" which makes the oversimplified assumption that all disturbances and failures are of equal probability. Instead, ENTSO-E, which is regulated by ACER, requires operators to use FCs to quantify the expected performance, by considering the uncertainty in weather conditions, for example, ENTSO-E (2023b). Based on this methodology, operators can prioritize actions for preparedness, rapid response, recovery, and adaptation.

Furthermore, analysts in insurance companies utilize representative fragility models within catastrophe (aka CAT) modeling and carry out Life Cycle Cost Assessment (LCCA), considering future climate scenarios, to provide operators with low cost-effective insurance schemes (Marsh 2023). Employing insurance as a means of transferring risk to absorb financial losses and expedite considerably the recovery process associated with climate-related risks represents an initial shift away from reactive crisis response and toward proactive risk management and anticipation. According to findings in a report from the insurance sector, a mere 1% rise in insurance coverage has the potential to curtail the burden on taxpayers and governments from global climate-related disasters by a substantial 22% (Lloyd's 2012). Finally, governmental authorities widely use benefit–cost analyses, which incorporate FCs, to decide if a project is cost-effective and allocate resources. For

---

**BOX 1** | A Risk Management Framework of Power Grid Assets Involving the Main Applications of FCs.

Including inter-sectoral actors in the risk management of power grid can improve the efficiency of risk mitigation measures. Firstly, risk analysts conduct multi-hazard identification and modeling, considering different sources of uncertainty related to structural parameters, climate change, and climatic projections. Furthermore, risk quantification is conducted using FCs, which enable reliability evaluation and stress testing, assigning a "pass" or "not pass" grade for each electrical asset. Critical assets and system reliability are identified by risk analysts, who report the outcomes to other actors.

Next, safety reports for power grids must include operational and environmental hazards. Power operators or stakeholders are obligated to evaluate the code compliance of assets and determine the need for risk treatment for non-compliant assets using FCs. Rapid tracking of the most vulnerable assets of a power grid is carried out using FCs, enabling the activation of emergency response. Furthermore, by monitoring impacts it becomes possible to modify FCs according to the nature of the ongoing climate change. Lessons learned from previous incidents can inform inspection prioritization for recovery and climate change adaptation measures.

Additionally, insurance companies incorporate FCs into their software for rapid risk assessment or life-cycle costing. The accuracy of the risk calculations they communicate to their customers, as well as the insurance schemes they provide, rely on the FCs. Hence, the use of robust fragility models is of utmost importance to avoid liability issues. In the aftermath of events, insurance companies handle insurance coverage.

Finally, policy makers such as governmental bodies in the ministry of energy, or local authorities communicate with other actors and prioritize investments—accounting for social dimensions of climate change—to develop and implement policies, aimed at achieving sustainability at the community level. The effectiveness of investments relies on the accuracy of cost–benefit analyses, which employ FCs.

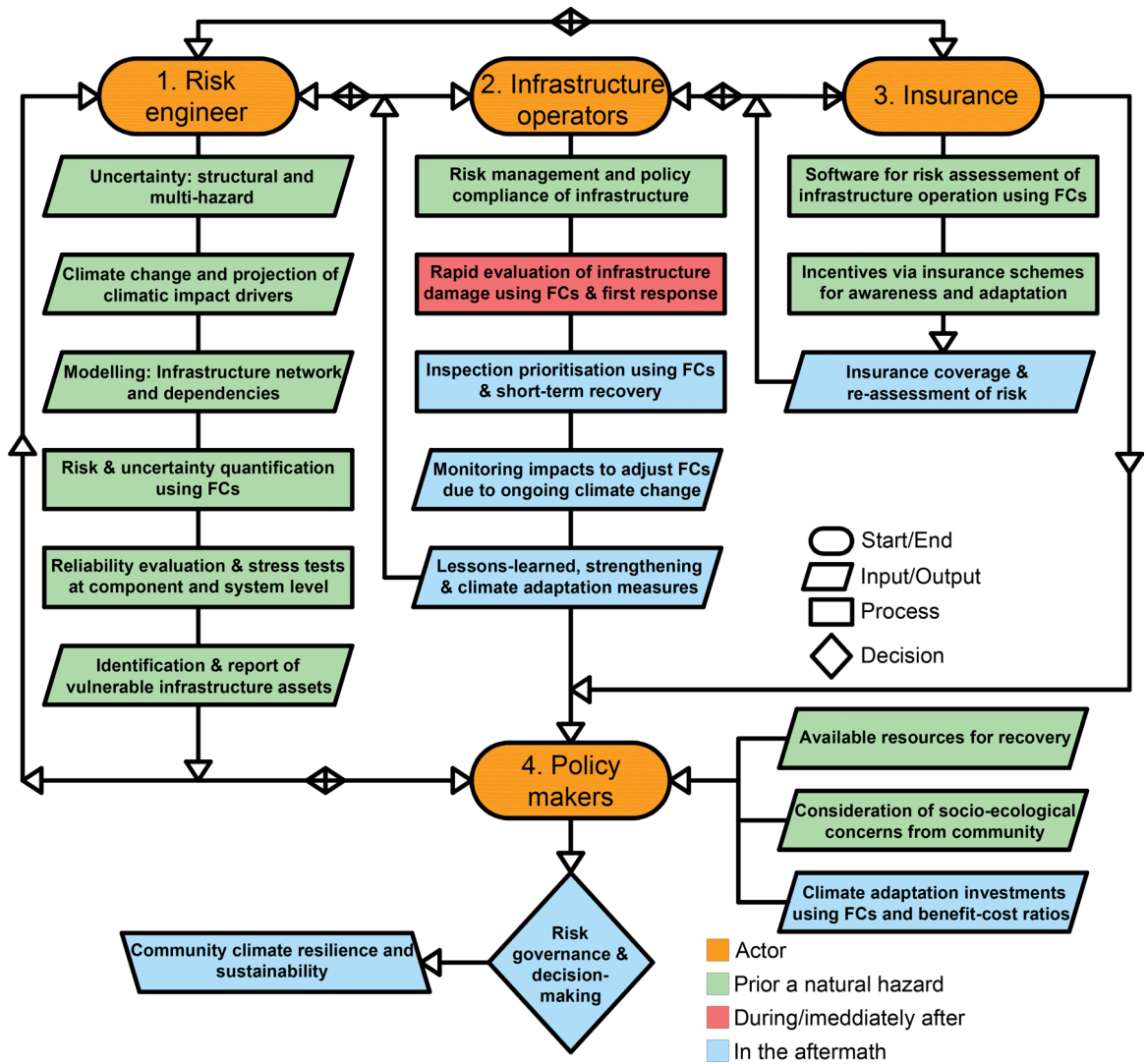

**FIGURE 8** | A flowchart with the main applications of fragility curves (FCs) and decision-making processes in the risk management of power grid assets by different actors.

example, FEMA requires the completion of a benefit–cost-ratio worksheet when applying for Federal disaster funds (Kabre and Weimar 2022). Also, ENTSO-E (2023a) requires projects to evaluate cost–benefit ratios of adaptation measures, by taking into account several climatic conditions (the so-called "climate year").

## 5.2 | Climate Change Adaptation Measures

Table 3 summarizes various climate change adaptation strategies for power grid assets, focusing on measures tailored to specific climate hazards. The Sixth Assessment Report (AR6) of the Intergovernmental Panel on Climate Change (IPCC; IPCC 2021) highlights that climatic-impact drivers (physical climate system conditions such as mean or extremes values of temperature, rainfall, snowfall, windstorm, and floods) are expected to increase in certain regions (IPCC 2021). These climate drivers, discussed also in IPCC (IPCC 2022), are critical considerations in the design, assessment, and management of power grid assets to enhance climate resilience.

Common adaptation measures are applicable across different climate hazards. For example, structural strengthening, re-design accounting for multiple-hazard effects and climatic projections as well as frequent inspection of assets and use of underground cables (Najafi Tari, Sepasian, and Tourandaz Kenari 2021). Also, land-use planning to find alternative power grid paths in hazard-prone regions can considerably reduce the risk and enhance resilience. Vegetation management, which includes trimming, removing hazard trees and other airborne objects near overhead power lines to minimize interference risks, was identified as the most effective option for climate resilience enhancement by Hou et al. (2023). In view of climate change effects, this adaptation approach can reduce the risk against windstorms, heatwaves, and wildfires. Additionally, an effective communication channel between inter-sectoral actors, for example, power grid operators and meteorological agencies, as highlighted in Figure 8, can improve risk management considerably.

Cost–benefit analyses can improve decision-making for climate change adaptation investments against natural hazards

**TABLE 3** | Risk mitigation and climate change adaptation measures of power grid assets against climate hazards.

| Climate hazard | Adaptation measures |
| --- | --- |
| Windstorm | • Risk assessment models with multiple hazard scenarios and climatic projections for possible upgrading or redesigning of towers (Snaiki and Parida 2023; Yuan et al. 2018)<br>• Consideration of mechanical fuses to reduce conductor breakages (Eurelectric 2022)<br>• Land-use planning for alternative power grid paths (IPCC 2022)<br>• Communication channel between operators and meteorological agencies<br>• Frequent inspection and maintenance<br>• Cutting vegetation and removing airborne debris near critical power grid assets (Hou et al. 2023)<br>• Use of underground electrical cables in specific regions (Najafi Tari, Sepasian, and Tourandaz Kenari 2021)<br>• Cost–benefit analysis for efficient decision-making, for example, replacement of towers and pylons with another type (T-pylons or steel poles), placement below ground, or relocation (Dehghani, Mohammadi, and Karimi 2023)<br>• Machine-learning approaches for uncertainty reduction and efficient decision-making (Noebels, Preece, and Panteli 2022)<br>• Consideration of aging effects and foundation scour |
| Lightning strikes | • Risk assessment models that combine exposure, failure rates, and climatic projections (Souto, Taylor, and Wilkinson 2022)<br>• Shield wires, surge arresters, lightning masts (IEEE 1410 2011)<br>• High-quality insulator, for example, polymers or fiberglass<br>• Vector shift protection<br>• Load shedding with real-time load assessment of feeders using telecommunications (Bialek 2020)<br>• Load shedding at lower voltage level<br>• Grid reconfiguration that reduces the probability of power outages |
| Heat waves and wildfires | • Proactive outage prediction modeling and shutoff due to wildfire progression (Dian et al. 2019)<br>• Proactive generation redispatch<br>• Hardening, maintenance planning, and vegetation management Hou et al. (2023)<br>• Installation of solar panels and mobile energy storage systems to increase redundancy, accounting for weather uncertainties (Moreno et al. 2022)<br>• Network reconfiguration (Abdelmalak and Benidris 2022)<br>• Microgrid formation<br>• Participation of system operators in decision-making |
| Flood | • Flood barriers; portable flood defense measures; earth bunds; flood doors and gates; drainage systems and pumping stations; flood storage reservoirs; and land management-based measures (Karagiannis, Chondrogiannis, et al. 2017)<br>• Locating the substation above flood levels, elevating the equipment, or waterproofing<br>• Levee protection and use of nature-based solutions (Sayers et al. 2020)<br>• Early warning for shutting down flood-prone substations, activating surge mechanisms, and staging repair capabilities at the edge of flood zone (Sayers et al. 2020) |

(Dehghani, Mohammadi, and Karimi 2023). FCs that account for aging effects are integrated with the hazard curve to derive the expected failure probability, for example, tower or pylon in the event of a natural disaster. This probability is then multiplied by the asset's repair cost over its lifetime, and the total repair cost for all assets is calculated by summing these products (Francis et al. 2011). For example, in the case of windstorms, Salman and Li (2016) observed steel poles offer higher long-term structural reliability than wood poles, wood poles are more cost-efficient for up to a 50-years, after which decay increases replacement costs. Also, Shafieezadeh et al. (2014) and Darestani and Shafieezadeh (2019) noted significant discrepancies in the fragility level among wood poles aged over 25 years, suggesting that risk-targeted inspections can yield annual savings. Cost–benefit analyses for transmission towers are less common, though guidelines by ENTSO-E (2023a) recommend estimating and reporting environmental-related costs, for example, for maintenance or retrofitting.

Furthermore, probabilistic risk assessment models can play a crucial role in reducing power outages, and can also aid in adaptation efforts. For example, these models can combine exposure and failure rates with weather forecasts, in partnership with meteorological offices, to account for increased lightning activity due to climate change (Souto, Taylor, and Wilkinson 2022). They can guide revisions in design criteria, improve the maintenance of lightning protection equipment, and suggest the use of shield wires and lightning arresters. Also, power system planning, such as grid reconfiguration, voltage-based load shedding, and real-time load assessment of feeders with advanced telecommunications can further reduce lightning-related outages (Bialek 2020).

In response to the heightened risks of extreme weather events, particularly heat wave-induced disturbances, several adaptation strategies have been proposed. These include developing proactive line outage prediction models that simulate wildfire spread (Dian et al. 2019) and power generation redispatch strategies that address the spatiotemporal aspects of wildfires to minimize load curtailment and costs. Other measures involve establishing microgrids to enhance grid resilience, increasing the use of renewable energy for redundancy, and reconfiguring networks to account for climate change (Abdelmalak and Benidris 2022). Furthermore, the repairing of substations typically costs between 86%–88% of the total repair costs, while restoring damaged towers and economic disruptions costs between 8%–12% and 2%–4%, respectively (Karagiannis, Turksezer, et al. 2017). Hence, attention should be given to the protection of indoor or outdoor substations. Adaptation measures for flooding, whether fluvial, pluvial, or coastal, include relocating or protecting substations with barriers, bollards, drainage systems, and flood storage reservoirs (Sayers et al. 2020). Other measures include levees, early warning systems, and staged repair capabilities to reduce flood risk and mitigate consequences (Karagiannis, Turksezer, et al. 2017).

Finally, the accuracy of probabilistic risk assessment models and decision-making for adaptation measures can be enhanced with the use of machine-learning techniques. Noebels, Preece, and Panteli (2022) and Venkatasubramanian et al. (2023) proposed a risk-assessment tool for finding optimal preventive measures in real-time based on Monte Carlo simulations trained on environmental, topographical, and grid-related data. Similar studies that employ various machine-learning approaches and metrics for decision-making can be found in Duchesne et al. (2020) and Lu and Zhang (2022). In fact, machine-learning techniques can take advantage of different data sources such as the contributing factors presented in Table 2 to reduce the uncertainty of risk models. Such an application can be found in Cerrai, Watson, and Anagnostou (2019) for vegetation management.

## 6 | Conclusion

This article has highlighted the most recent developments and gaps in the fragility and climate change adaptation of power grid assets. Built upon the findings, the following conclusions can be drawn:

- Prolonged periods of power shortage and significant financial losses are observed for major incidents due to the lack of robust risk assessment models and preparedness. In the case of transmission and distribution networks, the most common causes and contributing factors are due to wind loading above design level, falling trees and debris, and aging effects. For substations, exceedance of inundation level, poor maintenance of equipment, and lack of protection barriers can result in damage to equipment and short circuits.
- New fragility models are needed to enhance the resilience assessment of power grid assets against wind hazard, flood, and wildfire, accounting for climate change projections. For example, there are no analytical wind and multi-hazard models available to encompass the diverse range of transmission tower types across different ecosystems. A similar conclusion can be made with respect to substations and flood hazard. Wildfire was identified as a significant hazard, but there is still limited research on the topic. Also, the impact of diverse non-stationary climate scenarios on these hazards has not been thoroughly investigated. These scenarios can cause variations for the aforementioned hazards, which can vary in intensity and frequency based on the geographical area.
- A risk management framework highlights the inter-sectoral application of fragility curves, and the need to form proper communication channels among risk analysts, power operators, insurance companies, and policymakers to foster resilience and sustainability in the energy sector. This framework can be considered as an adaptation measure per se.
- Finally, climate change adaptation measures need to be considered to enhance the resilience of assets. The impact of climate-aware structural strengthening, wind or flood barriers, and mechanical fuses on conductors can be quantified through fragility models. Also, fragility models can inform decision-making regarding the relocation or extension of a power network in climate hazard-prone areas, as well as emergency response planning. Proactive measures such as power outage management and underground cables can be taken with the exploitation of fragility curves. Cost–benefit analyses can be used to evaluate the best trade-offs among these investments. The cost–benefit of these investments has not been examined sufficiently with the use of FCs for power grid assets, especially transmission networks. In a more holistic approach, climate adaptation measures can be integrated with strategies against anthropogenic attacks, such as conflicts (Mitoulis et al. 2023) or cyber threats (Linkov et al. 2013; Pöyhönen 2022), thereby improving overall infrastructure resilience.

**Author Contributions**

**George Karagiannakis:** conceptualization (equal), data curation (lead), formal analysis (lead), investigation (lead), methodology (lead), resources (lead), validation (lead), visualization (lead), writing – original draft (lead), writing – review and editing (lead). **Mathaios Panteli:** conceptualization (equal), methodology (equal), supervision (supporting), writing – original draft (supporting), writing – review and editing (equal). **Sotirios Argyroudis:** conceptualization (equal), funding acquisition (lead), methodology (lead), project administration (lead), supervision (lead), writing – original draft (supporting), writing – review and editing (lead).

**Acknowledgments**

The authors extend their sincere gratitude to the Editor-in-Chief, Domain Editor, and the two anonymous Reviewers for their insightful comments and constructive suggestions, which significantly enhanced the quality of this article. We also wish to thank Liane Frydland from Brunel University London for her meticulous proofreading of the manuscript.

**Conflicts of Interest**

The authors declare no conflicts of interest.

**Data Availability Statement**

Data sharing is not applicable to this article as no new data were created or analyzed in this study.

**Further Reading**

For climate change adaptation measures in the power grid sector, see: https://www-pub.iaea.org/MTCD/Publications/PDF/P1847_web.pdf; https://resilience.eurelectric.org/wp-content/uploads/2022/12/The-Coming-Storm_011222.pdf; https://www.eea.europa.eu/publications/adaptation-in-energy-system.

For the concept of resilience and metrics in power grid, see: https://www.sciencedirect.com/science/article/pii/S1364032123007207; https://ieeexplore.ieee.org/abstract/document/7091066; https://ieeexplore.ieee.org/abstract/document/7842605.

EU policy for climate change adaptation and cost–benefit analysis can be found in: https://climate.ec.europa.eu/eu-action/adaptation-climate-change/eu-adaptation-strategy_en; https://eur-lex.europa.eu/legal-content/EN/TXT/PDF/?uri=CELEX:32022R0869&qid=1692953203005.

**Related WIREs Articles**

A compound event framework for understanding extreme impacts

Vulnerability and resilience of power systems infrastructure to natural hazards and climate change

Implications of climate change for railway infrastructure

**References**

Abdelmalak, M., and M. Benidris. 2022. "Enhancing Power System Operational Resilience Against Wildfires." *IEEE Transactions on Industry Applications* 58: 1611–1621. https://doi.org/10.1109/TIA.2022.3145765.

ACER. 2019. "Methodology for coordinating operational security analysis." In *ACER Decision on CSAM: Annex I* (text rectified by corrigendum of 27 January 2023). https://www.acer.europa.eu/sites/default/files/documents/IndividualDecisions_annex/ACER_Decision_CSAM-AnnexI_Rectified.pdf.

Alipour, A., and S. Dikshit. 2023. "A Moment-Matching Method for Fragility Analysis of Transmission Towers Under Straight Line Winds." *Reliability Engineering & System Safety* 236: 109241. https://doi.org/10.1016/j.ress.2023.109241.

ANSI. 2015. *Wood Poles Specifications and Dimensions*. Washington, DC: Accredited Standards Committee.

Argyroudis, S. A. 2022. "Resilience Metrics for Transport Networks: A Review and Practical Examples for Bridges." *Proceedings of the Institution of Civil Engineers: Bridge Engineering* 175, no. 3: 179–192. https://doi.org/10.1680/jbren.21.00075.

Argyroudis, S. A., S. Fotopoulou, S. Karafagka, et al. 2020. "A Risk-Based Multi-Level Stress Test Methodology: Application to Six Critical Non-nuclear Infrastructures in Europe." *Natural Hazards* 100, no. 2: 595–633. https://doi.org/10.1007/s11069-019-03828-5.

Argyroudis, S. A., S. Mitoulis, M. G. Winter, and A. M. Kaynia. 2019. "Fragility of Transport Assets Exposed to Multiple Hazards: State-of-the-Art Review Toward Infrastructural Resilience." *Reliability Engineering and System Safety* 191: 106567. https://doi.org/10.1016/j.ress.2019.106567.

Argyroudis, S. A., and S. A. Mitoulis. 2021. "Vulnerability of Bridges to Individual and Multiple Hazards—Floods and Earthquakes." *Reliability Engineering and System Safety* 210: 107564. https://doi.org/10.1016/j.ress.2021.107564.

Argyroudis, S. A., S. A. Mitoulis, L. Hofer, M. A. Zanini, E. Tubaldi, and D. M. Frangopol. 2020. "Resilience Assessment Framework for Critical Infrastructure in a Multi-Hazard Environment: Case Study on Transport Assets." *Science of the Total Environment* 714: 136854. https://doi.org/10.1016/j.scitotenv.2020.136854.

ASCE. 2013. *Minimum Design Loads for Buildings and Other Structures*. Reston, VA: ASCE Standard. https://doi.org/10.1061/9780784412916.

Bakalis, K., and D. Vamvatsikos. 2018. "Seismic Fragility Functions via Nonlinear Response History Analysis." *Journal of Structural Engineering (United States)* 144, no. 10: 04018181. https://doi.org/10.1061/(ASCE)ST.1943-541X.0002141.

Bao, J., X. Wang, Y. Zheng, et al. 2021. "Resilience-Oriented Transmission Line Fragility Modeling and Real-Time Risk Assessment of Thunderstorms." *IEEE Transactions on Power Delivery* 36: 2363–2373. https://doi.org/10.1109/TPWRD.2021.3066157.

BBC. 2020. *Storm Ciara: In Pictures*. London, UK: British Broadcasting Corporation. https://www.bbc.com/news/uk-51436720.

Bi, W., L. Tian, C. Li, and Z. Ma. 2023. "A Kriging-Based Probabilistic Framework for Multi-Hazard Performance Assessment of Transmission Tower-Line Systems Under Coupled Wind and Rain Loads." *Reliability Engineering & System Safety* 240: 109615. https://doi.org/10.1016/j.ress.2023.109615.

Bialek, J. 2020. "What Does the GB Power Outage on 9 August 2019 Tell Us About the Current State of Decarbonised Power Systems?" *Energy Policy* 146: 111821. https://doi.org/10.1016/j.enpol.2020.111821.

Bjarnadottir, S., Y. Li, and M. G. Stewart. 2013. "Hurricane Risk Assessment of Power Distribution Poles Considering Impacts of a Changing Climate." *Journal of Infrastructure Systems* 19: 12–24. https://doi.org/10.1061/(asce)is.1943-555x.0000108.

Brown, R. 2009. "Cost–Benefit Analysis of the Deployment of Utility Infrastructure Upgrades and Storm Hardening Programs." Final Report, Project No. 36375, Quanta Technology. https://ftp.puc.texas.gov/public/puct-info/industry/electric/reports/infra/utlity_infrastructure_upgrades_rpt.pdf.

Cai, Y., and J. Wan. 2021. "Wind-Resistant Capacity Modeling for Electric Transmission Line Towers Using Kriging Surrogates and Its Application to Structural Fragility." *Applied Sciences (Switzerland)* 11, no. 11: 4714. https://doi.org/10.3390/app11114714.

Cai, Y., Q. Xie, S. Xue, L. Hu, and A. Kareem. 2019. "Fragility Modelling Framework for Transmission Line Towers Under Winds." *Engineering Structures* 191: 686–697. https://doi.org/10.1016/j.engstruct.2019.04.096.

Cao, J., Y. Du, Y. Ding, et al. 2022. "Comprehensive Assessment of Lightning Protection Schemes for 10 kV Overhead Distribution Lines." *IEEE Transactions on Power Delivery* 37, no. 3: 2326–2336. https://doi.org/10.1109/TPWRD.2021.3110248.

Cavalieri, F., P. Franchin, and P. E. Pinto. 2014. *Fragility Functions of Electric Power Stations*, 157–185. Dordrecht, Netherlands: Springer. https://doi.org/10.1007/978-94-007-7872-6_6.

CER Directive. 2022. Directive of the European Parliament and of the Council of 14 December 2022 on the Resilience of Critical Entities and Repealing Council Directive 2008/114/EC.

Cerrai, D., P. Watson, and E. N. Anagnostou. 2019. "Assessing the Effects of a Vegetation Management Standard on Distribution Grid Outage Rates." *Electric Power Systems Research* 175: 105909. https://doi.org/10.1016/j.epsr.2019.105909.

Chen, L., C. Li, Z. Xin, and S. Nie. 2022. "Simulation and Risk Assessment of Power System With Cascading Faults Caused by Strong Wind Weather." *International Journal of Electrical Power & Energy Systems* 143: 108462. https://doi.org/10.1016/j.ijepes.2022.108462.

Cherp, A. 2012. "Defining Energy Security Takes More Than Asking Around." *Energy Policy* 48: 841–842. https://doi.org/10.1016/j.enpol.2012.02.016.

Cimellaro, G. P., A. M. Reinhorn, M. Bruneau, and A. Rutenberg. 2006. *Multi Dimensional Fragility of Structures: Formulation and Evaluation.* Report number: MCEER-06-0002, Multidisciplinary Center for Earthquake Engineering Research, University at Buffalo - the State University of New York. https://www.eng.buffalo.edu/mceer-reports/06/06-0002.pdf.

City-Journal. 2019. California Goes Dark. https://www.city-journal.org/article/california-goes-dark.

CNN. 2023. Πτώση δυο πυλώνων στη Θεσσαλία από τη διάβρωση του εδάφους—Κανένα πρόβλημα στο δίκτο μεταφοράς (in Greek). https://www.cnn.gr/ellada/story/381344/ptosi-dyo-pylonon-sti-thessalia-apo-ti-diavrosi-tou-edafous-kanena-provlima-sto-dikto-metaforas.

Darestani, Y. M., and A. Shafieezadeh. 2019. "Multi-Dimensional Wind Fragility Functions for Wood Utility Poles." *Engineering Structures* 183: 937–948. https://doi.org/10.1016/j.engstruct.2019.01.048.

Dehghani, F., M. Mohammadi, and M. Karimi. 2023. "Age-Dependent Resilience Assessment and Quantification of Distribution Systems Under Extreme Weather Events." *International Journal of Electrical Power & Energy Systems* 150: 109089. https://doi.org/10.1016/j.ijepes.2023.109089.

DeMenno, M. B., R. J. Broderick, and R. F. Jeffers. 2022. "From Systemic Financial Risk to Grid Resilience: Embedding Stress Testing in Electric Utility Investment Strategies and Regulatory Processes." *Sustainable and Resilient Infrastructure* 7, no. 6: 673–694. https://doi.org/10.1080/23789689.2021.2015833.

Di Sarno, L., and G. Karagiannakis. 2020. "On the Seismic Fragility of Pipe Racks—Piping Systems Considering Soil-Structure Interaction." *Bulletin of Earthquake Engineering* 18, no. 6: 2723–2757. https://doi.org/10.1007/s10518-020-00797-0.

Dian, S., P. Cheng, Q. Ye, et al. 2019. "Integrating Wildfires Propagation Prediction Into Early Warning of Electrical Transmission Line Outages." *IEEE Access* 7: 27586–27603. https://doi.org/10.1109/ACCESS.2019.2894141.

Dimensions. 2018. Database [Free]. https://app.dimensions.ai.

DOE. 2016. U.S. Department of Energy, Office of Energy Policy and Systems Analysis, Climate Change and the Electricity Sector: Guide for Climate Change Resilience Planning. https://www.energy.gov/sites/prod/files/2016/10/f33/ClimateChangeandtheElectricitySectorGuideforClimateChangeResiliencePlanningSeptember2016_0.pdf.

Dong, K., Q. Shen, C. Wang, Y. Dong, and Z. Lu. 2022. "Research on Online Monitoring and Early Warning System of Transmission Line Galloping Based on Multi-Source Data." *Journal of Measurements in Engineering* 10, no. 4: 188–198. https://doi.org/10.21595/jme.2022.22818.

Dos Reis, F. B., P. Royer, V. H. Chalishazar, et al. 2022. "Methodology to Calibrate Fragility Curves Using Limited Real-World Data." *IEEE Power and Energy Society General Meeting (PESGM)*: 1–5. https://doi.org/10.1109/PESGM48719.2022.9916809.

Duchesne, L., E. Karangelos, A. Sutera, and L. Wehenkel. 2020. "Machine Learning for Ranking Day-Ahead Decisions in the Context of Short-Term Operation Planning." *Electric Power Systems Research* 189: 106548. https://doi.org/10.1016/j.epsr.2020.106548.

Dumas, M., B. Kc, and C. I. Cunliff. 2019. *Extreme Weather and Climate Vulnerabilities of the Electric Grid: A Summary of Environmental Sensitivity Quantification Methods.* Oak Ridge, UK: OAK Ridge National Laboratory.

Dunn, S., S. Wilkinson, D. Alderson, H. Fowler, and C. Galasso. 2018. "Fragility Curves for Assessing the Resilience of Electricity Networks Constructed From an Extensive Fault Database." *Natural Hazards Review* 19, no. 1: 1–10. https://doi.org/10.1061/(asce)nh.1527-6996.0000267.

EC. 2018. *Study on the Quality of Electricity Market Data of Transmission System Operators, Electricity Supply Disruptions, and Their Impact on the European Electricity Markets.* Brussels, Belgium: European Commission. https://energy.ec.europa.eu/study-quality-electricitymar.

EC. 2021. Communication From the Commission to the European Parliament, The Council, The European Economic and Social Committee and The Committee of the Regions Forging a climate-resilient Europe—The New EU Strategy on Adaptation to Climate Change.

Eidinger, J. 2011. *2010–2011 Christchurch Earthquakes, Impact to Electric Power Systems.* Olympic Valley, CA: G&E Engineering Systems. http://www.geengineeringsystems.com/ewExternalFiles/ChristchurchPower11.2.11lr.pdf.

Eidinger, J. M., and L. Kempner. 2012. "Reliability of Transmission Towers Under Extreme Wind and Ice Loading." In *44th International Conference on Large High Voltage Electric Systems*.

EMBER. 2023. Report of the European Electricity Transition in Europe in 2022. https://ember-climate.org/insights/research/european-electricity-review-2023/.

EN. 2005. 1991-1-4, Eurocode 1: Actions on structures—Part 1–4: General actions—Wind actions [Authority: The European Union Per Regulation 305/2011, Directive 98/34/EC, Directive 2004/18/EC].

ENTSO-E. 2013. "European Network of Transmission System Operators for Electricity." In *Nordic Grid Disturbance Statistics 2012*, 1–92. Brussels, Belgium: ENTSO-E AISBL.

ENTSO-E. 2022. *Incident Classification Scale Annual Report.* Brussels, Belgium: European Network of Transmission System Operators for Electricity. https://www.entsoe.eu/network_codes/sys-ops/annual-reports/.

ENTSO-E. 2023a. 4th Guideline for Cost Benefit Analysis of Grid Development Projects. https://tyndp.entsoe.eu/resources/4-th-entso-e-guideline-for-cost-benefit-analysis-of-grid-development-projects.

ENTSO-E. 2023b. Biennial Progress Report on Operational Probabilistic Coordinated Security Assessment and Risk Management. https://eepublicdownloads.blob.core.windows.net/public-cdn-container/clean-documents/SOCdocuments/SOCReports/entso-e_report_PRA_2023_231208_FINAL.pdf.

ENTSO-E. 2023c. European Network of Transmission System Operators for Electricity—Power Grid Map at Pan European Level. https://www.entsoe.eu/data/map/.

ENTSO-E. 2023d. Installed Capacity per Production Type in Europe. https://transparency.entsoe.eu/generation/r2/installedGenerationCapacityAggregation/show.

Eurelectric. 2022. *The Coming Storm: Building Electricity to Extreme Weather.* Brussels, Belgium: Union of the Electricity Industry. https://resilience.eurelectric.org/wp-content/uploads/2022/12/The-Coming-Storm_011222.pdf.

FEMA. 2013. *Multi-Hazard Loss Estimation Methodology: Flood Model.* Washington, DC: Department of Homeland Security Federal Emergency Management Agency Mitigation Division.

Feng, K., M. Ouyang, and N. Lin. 2022. "Tropical Cyclone-Blackout-Heatwave Compound Hazard Resilience in a Changing Climate." *Nature Communications* 13: 4421. https://doi.org/10.1038/s41467-022-32018-4.

Francis, R. A., S. M. Falconi, R. Nateghi, and S. D. Guikema. 2011. "Probabilistic Life Cycle Analysis Model for Evaluating Electric Power Infrastructure Risk Mitigation Investments." *Climatic Change* 106, no. 1: 31–55. https://doi.org/10.1007/s10584-010-0001-9.

Fu, X., and H.-N. Li. 2018. "Uncertainty Analysis of the Strength Capacity and Failure Path for a Transmission Tower Under a Wind Load." *Journal of Wind Engineering and Industrial Aerodynamics* 173: 147–155. https://doi.org/10.1016/j.jweia.2017.12.009.

Fu, X., H. N. Li, and G. Li. 2016. "Fragility Analysis and Estimation of Collapse Status for Transmission Tower Subjected to Wind and Rain Loads." *Structural Safety* 58: 1–10. https://doi.org/10.1016/j.strusafe.2015.08.002.

Fu, X., H. N. Li, G. Li, and Z. Q. Dong. 2020. "Fragility Analysis of a Transmission Tower Under Combined Wind and Rain Loads." *Journal of Wind Engineering and Industrial Aerodynamics* 199: 104098. https://doi.org/10.1016/j.jweia.2020.104098.

Fu, X., H.-N. Li, L. Tian, J. Wang, and H. Cheng. 2019. "Fragility Analysis of Transmission Line Subjected to Wind Loading." *Journal of Performance of Constructed Facilities* 33, no. 4: 04019044. https://doi.org/10.1061/(asce)cf.1943-5509.0001311.

Hallegatte, S., J. Rentschler, and J. Rozenberg. 2019. *Lifelines: The Resilient Infrastructure Opportunity. Sustainable Infrastructure Series*. Washington, DC: World Bank. https://openknowledge.worldbank.org/handle/10986/31805.

HAZUS. 2022a. *(Hazard U.S.). Hazus—MH 5.1, Earthquake Model, Technical Manual*. Washington, DC: FEMA.

HAZUS. 2022b. *(Hazard U.S.). Hazus—MH 5.1, Flood Model, Technical Manual*. Washington, DC: FEMA.

HAZUS. 2022c. *(Hazard U.S.). Hazus—MH 5.1, Hurricane Model, Technical Manual*. Washington, DC: FEMA.

Hofmann, M., G. H. Kjølle, and O. Gjerde. 2013. Vulnerability Indicators for Electric Power Grids. http://hdl.handle.net/11250/2598688.

Hou, G., and S. Chen. 2020. "Probabilistic Modeling of Disrupted Infrastructures Due to Fallen Trees Subjected to Extreme Winds in Urban Community." *Natural Hazards* 102: 1323–1350. https://doi.org/10.1007/s11069-020-03969-y.

Hou, G., and K. K. Muraleetharan. 2023. "Modeling the Resilience of Power Distribution Systems Subjected to Extreme Winds Considering Tree Failures: An Integrated Framework." *International Journal of Disaster Risk Science* 14, no. 2: 194–208. https://doi.org/10.1007/s13753-023-00478-x.

Hou, G., K. K. Muraleetharan, V. Panchalogaranjan, et al. 2023. "Resilience Assessment and Enhancement Evaluation of Power Distribution Systems Subjected to Ice Storms." *Reliability Engineering and System Safety* 230: 108964. https://doi.org/10.1016/j.ress.2022.108964.

Hughes, W., W. Zhang, D. Cerrai, A. Bagtzoglou, D. Wanik, and E. Anagnostou. 2022. "A Hybrid Physics-Based and Data-Driven Model for Power Distribution System Infrastructure Hardening and Outage Simulation." *Reliability Engineering & System Safety* 225: 108628. https://doi.org/10.1016/j.ress.2022.108628.

IEEE 1410. 2011. IEEE Guide for Improving the Lightning Performance of Electric Power Overhead Distribution Lines. In IEEE Std 1410-2010 (Revision of IEEE Std 1410-2004).

IPCC. 2021. "Climate Change 2021: The Physical Science Basis." In *Contribution of Working Group I to the Sixth Assessment Report of the Intergovernmental Panel on Climate Change*, edited by V. Masson-Delmotte, P. Zhai, A. Pirani, et al. Cambridge, UK: Cambridge University Press. https://doi.org/10.1017/9781009157896.

IPCC. 2022. "IPCC, 2022: Climate Change 2022: Impacts, Adaptation and Vulnerability." In *Contribution of Working Group II to the Sixth Assessment Report of the Intergovernmental Panel on Climate Change*, edited by H.-O. Pörtner, D. C. Roberts, M. Tignor, E. S. Poloczanska, and K. Mintenbe. Cambridge, UK: Cambridge University Press. https://doi.org/10.1017/9781009325844.

IPCC. 2023. Intergovernmental Panel on Climate Change 2023 – Summary for policy makers of the synthesis report, Issue 2, pages 2–5. https://www.ipcc.ch/report/ar6/syr/downloads/report/IPCC_AR6_SYR_SPM.pdf.

Jeddi, A. B., A. Shafieezadeh, J. Hur, J. G. Ha, D. Hahm, and M. K. Kim. 2022. "Multi-Hazard Typhoon and Earthquake Collapse Fragility Models for Transmission Towers: An Active Learning Reliability Approach Using Gradient Boosting Classifiers." *Earthquake Engineering and Structural Dynamics* 51: 3552–3573. https://doi.org/10.1002/eqe.3735.

Kabre, W., and M. R. Weimar. 2022. *Fragility Functions Resource Report Documented Sources for Electricity and Water Resilience Valuation, Prepared for the U.S. Department of Energy Under Contract DE-AC05-76RL01830*. Richland, WA: Pacific Northwest National Laboratory.

Karagiannakis, G., L. Di Sarno, A. Necci, and E. Krausmann. 2022. "Seismic Risk Assessment of Supporting Structures and Process Piping for Accident Prevention in Chemical Facilities." *International Journal of Disaster Risk Reduction* 69: 102748. https://doi.org/10.1016/j.ijdrr.2021.102748.

Karagiannakis, G., M. Panteli, and S. Argyroudis. 2023. "Fragility Assessment of Power Grid Infrastructure Towards Climate Resilience and Adaptation." In *63rd ESReDA Seminar on Resilience Assessment: Methodological Challenges and Applications to Critical Infrastructures*. Ispra, Italy: Joint Research Center. https://www.researchgate.net/publication/378487409_Fragility_assessment_of_power_grid_infrastructure_towards_climate_resilience_and_adaptation.

Karagiannis, G., M. Cardarilli, Z. Turksezer, et al. 2019. Climate Change and Critical Infrastructure—Storms (Issue February). https://doi.org/10.2760/986436.

Karagiannis, G. M., S. Chondrogiannis, E. K. Zehra, and I. Turksezer. 2017. "Power Grid Recovery After Natural Hazard Impact—A Science for Policy Report." In *Science for Policy Report by the Joint Research Centre (JRC)*. Luxembourg: European Union. https://data.europa.eu/doi/10.2760/87402.

Karagiannis, G. M., Z. Turksezer, L. Alfieri, L. Feyen, and E. Krausmann. 2017. *Climate Change and Critical Infrastructure—Floods*. EUR 28855 EN, JRC109015. Luxembourg: Publications Office of the European Union. https://doi.org/10.2760/007069.

Karimi, M., S. N. Ravadanegh, and M.-R. Haghifam. 2021. "A Study on Resilient and Cost-Based Design in Power Distribution Network Against Severe Hurricane." *International Journal of Critical Infrastructure Protection* 35: 100469. https://doi.org/10.1016/j.ijcip.2021.100469.

Kazantzi, A. K., S. Moutsianos, K. Bakalis, and S.-A. Mitoulis. 2024. "Cause-Agnostic Bridge Damage State Identification Utilising Machine Learning." *Engineering Structures* 320: 118887. https://doi.org/10.1016/j.engstruct.2024.118887.

Kennedy, R. P., and M. K. Ravindra. 1984. "Seismic Fragilities for Nuclear Power Plant Risk Studies." *Nuclear Engineering and Design* 79: 47–68. https://doi.org/10.1016/0029-5493(84)90188-2.

Klinger, C., M. Mehdianpour, D. Klingbeil, D. Bettge, R. Häcker, and W. Baer. 2011. "Failure Analysis on Collapsed Towers of Overhead Electrical Lines in the Region Münsterland (Germany) 2005." *Engineering Failure Analysis* 18: 1873–1883. https://doi.org/10.1016/j.engfailanal.2011.07.004.

Koks, E. E., K. C. H. Van Ginkel, M. J. E. Van Marle, and A. Lemnitzer. 2022. "Brief Communication: Critical Infrastructure Impacts of the 2021 Mid-July Western European Flood Event." *Natural Hazards and Earth System Sciences* 22, no. 12: 3831–3838. https://doi.org/10.5194/nhess-22-3831-2022.

Koutsourelakis, P. S. 2010. "Assessing Structural Vulnerability Against Earthquakes Using Multi-Dimensional Fragility Surfaces: A Bayesian Framework." *Probabilistic Engineering Mechanics* 25: 49–60. https://doi.org/10.1016/j.probengmech.2009.05.005.

Krausmann, E., E. Renni, M. Campedel, and V. Cozzani. 2011. "Industrial Accidents Triggered by Earthquakes, Floods and Lightning: Lessons Learned From a Database Analysis." *Natural Hazards* 59: 285–300. https://doi.org/10.1007/s11069-011-9754-3.

KSU. 2021. Kansas State University. https://www.k-state.edu/research/about/seek/spring-2021/engineers-make-electric-grid-smarter-safer.html.

Lam, J. C., B. T. Adey, M. Heitzler, et al. 2018. "Stress Tests for a Road Network Using Fragility Functions and Functional Capacity Loss Functions." *Reliability Engineering & System Safety* 173: 78–93. https://doi.org/10.1016/j.ress.2018.01.015.

Lee, S., and Y. Ham. 2021. "Probabilistic Framework for Assessing the Vulnerability of Power Distribution Infrastructures Under Extreme Wind Conditions." *Sustainable Cities and Society* 65: 102587. https://doi.org/10.1016/j.scs.2020.102587.

Leonard, M., S. Westra, A. Phatak, et al. 2014. "A Compound Event Framework for Understanding Extreme Impacts." *WIREs Climate Change* 5, no. 1: 113–128. https://doi.org/10.1002/wcc.252.

Li, C., H. Pan, L. Tian, and W. Bi. 2022. "Lifetime Multi-Hazard Fragility Analysis of Transmission Towers Under Earthquake and Wind Considering Wind-Induced Fatigue Effect." *Structural Safety* 99: 102266. https://doi.org/10.1016/j.strusafe.2022.102266.

Li, Q., H. Jia, Q. Qiu, et al. 2022. "Typhoon-Induced Fragility Analysis of Transmission Tower in Ningbo Area Considering the Effect of Long-Term Corrosion." *Applied Sciences* 12, no. 9: 4774. https://doi.org/10.3390/app12094774.

Li, W., L. S. Cunningham, D. M. Schultz, S. Mander, C. K. Gan, and M. Panteli. 2024. "Structural Resilience of Pole-Mounted Substations Subjected to Flooding: Generalized Framework and a Malaysian Case Study." *ASCE-ASME Journal of Risk and Uncertainty in Engineering Systems, Part A: Civil Engineering* 10, no. 2: 04024008. https://doi.org/10.1061/AJRUA6.RUENG-1143.

Li, Y., A. Ahuja, and J. E. Padgett. 2012. "Review of Methods to Assess, Design for, and Mitigate Multiple Hazards." *Journal of Performance of Constructed Facilities* 26: 104–117. https://doi.org/10.1061/(asce)cf.1943-5509.0000279.

Liang, S., L. Zou, D. Wang, and H. Cao. 2015. "Investigation on Wind Tunnel Tests of a Full Aeroelastic Model of Electrical Transmission Tower-Line System." *Engineering Structures* 85: 63–72. https://doi.org/10.1016/j.engstruct.2014.11.042.

Linkov, I., D. A. Eisenberg, K. Plourde, T. P. Seager, J. Allen, and A. Kott. 2013. "Resilience Metrics for Cyber Systems." *Environment Systems and Decisions* 33, no. 4: 471–476. https://doi.org/10.1007/s10669-013-9485-y.

Linkov, I., B. D. Trump, J. Trump, et al. 2022. "Resilience Stress Testing for Critical Infrastructure." *International Journal of Disaster Risk Reduction* 82: 103323. https://doi.org/10.1016/j.ijdrr.2022.103323.

Liu, C., and Z. Yan. 2022. "Fragility Analysis of Wind-Induced Collapse of a Transmission Tower Considering Corrosion." *Buildings* 12, no. 10: 1500. https://doi.org/10.3390/buildings12101500.

Liu, X., K. Hou, H. Jia, et al. 2020. "A Planning-Oriented Resilience Assessment Framework for Transmission Systems Under Typhoon Disasters." *IEEE Transactions on Smart Grid* 11: 5431–5441. https://doi.org/10.1109/TSG.2020.3008228.

Liu, Y., S. Lei, and Y. Hou. 2019. "Overhead Transmission Line Outage Rate Estimation Under Wind Storms." *IEEJ Transactions on Electrical and Electronic Engineering* 14, no. 1: 57–66. https://doi.org/10.1002/tee.22765.

Lloyd. 2012. Global Underinsurance Report. https://assets.lloyds.com/assets/pdf-global-underinsurance-report-global-underinsurance-report/1/pdf-global-underinsurance-report-global-underinsurance-report.pdf.

Lu, J., J. Guo, Z. Jian, Y. Yang, and W. Tang. 2018. "Resilience Assessment and Its Enhancement in Tackling Adverse Impact of Ice Disasters for Power Transmission Systems." *Energies* 11, no. 9: 2272. https://doi.org/10.3390/en11092272.

Lu, Q., and W. Zhang. 2022. "Integrating Dynamic Bayesian Network and Physics-Based Modeling for Risk Analysis of a Time-Dependent Power Distribution System During Hurricanes." *Reliability Engineering and System Safety* 220: 108290. https://doi.org/10.1016/j.ress.2021.108290.

Ma, L., P. Bocchini, and V. Christou. 2020. "Fragility Models of Electrical Conductors in Power Transmission Networks Subjected to Hurricanes." *Structural Safety* 82: 101890. https://doi.org/10.1016/j.strusafe.2019.101890.

Ma, L., V. Christou, and P. Bocchini. 2022. "Framework for Probabilistic Simulation of Power Transmission Network Performance Under Hurricanes." *Reliability Engineering and System Safety* 217: 108072. https://doi.org/10.1016/j.ress.2021.108072.

Ma, L., M. Khazaali, and P. Bocchini. 2021. "Component-Based Fragility Analysis of Transmission Towers Subjected to Hurricane Wind Load." *Engineering Structures* 242, no. June: 112586. https://doi.org/10.1016/j.engstruct.2021.112586.

Ma, Y., Q. Dai, and W. Pang. 2020. "Reliability Assessment of Electrical Grids Subjected to Wind Hazards and Ice Accretion With Concurrent Wind." *Journal of Structural Engineering* 146, no. 7: 04020134. https://doi.org/10.1061/(asce)st.1943-541x.0002684.

Markolf, S. A., M. V. Chester, D. A. Eisenberg, et al. 2018. "Interdependent Infrastructure as Linked Social, Ecological, and Technological Systems (SETSs) to Address Lock-In and Enhance Resilience." *Earth's Future* 6: 1638–1659. https://doi.org/10.1029/2018EF000926.

Maroni, A., E. Tubaldi, H. McDonald, and D. Zonta. 2022. "A Monitoring-Based Classification System for Risk Management of Bridge Scour." *Proceedings of the Institution of Civil Engineers: Smart Infrastructure and Construction* 175, no. 2: 92–102. https://doi.org/10.1680/jsmic.21.00016.

Marsh. 2023. CAT Modeling: Which is the Best Approach for Your Project?, Energy and Power Sector. https://www.marsh.com/us/industries/energy-and-power/insights/cat-modeling-best-approach-project.html.

McKenna, G., S. A. Argyroudis, M. G. Winter, and S. A. Mitoulis. 2021. "Multiple Hazard Fragility Analysis for Granular Highway Embankments: Moisture Ingress and Scour." *Transportation Geotechnics* 26: 100431. https://doi.org/10.1016/j.trgeo.2020.100431.

Mitoulis, S.-A., S. Argyroudis, M. Panteli, et al. 2023. "Conflict-Resilience Framework for Critical Infrastructure Peacebuilding." *Sustainable Cities and Society* 91: 104405. https://doi.org/10.1016/j.scs.2023.104405.

Moreno, R., D. N. Trakas, M. Jamieson, et al. 2022. "Microgrids Against Wildfires: Distributed Energy Resources Enhance System Resilience." *IEEE Power and Energy Magazine* 20, no. 1: 78–89. https://doi.org/10.1109/MPE.2021.3122772.

Najafi Tari, A., M. S. Sepasian, and M. Tourandaz Kenari. 2021. "Resilience Assessment and Improvement of Distribution Networks Against Extreme Weather Events." *International Journal of Electrical Power & Energy Systems* 125: 106414. https://doi.org/10.1016/j.ijepes.2020.106414.

National Grid Electricity Transmission Plc. 2016. *Climate Change Adaptation, Second Round Adaptation Response*, 29. Warwick, UK: National Grid House. https://assets.publishing.service.gov.uk/media/5a80d3d7ed915d74e33fca2a/climate-adrep-national-grid.pdf.

Nazaripouya, H. 2020. "Power Grid Resilience Under Wildfire: A Review on Challenges and Solutions." In *IEEE Power and Energy Society General Meeting (PESGM)*, 1–5. Montreal, QC: Institute of Electrical and Electronics Engineers. https://doi.org/10.1109/PESGM41954.2020.9281708.

Nazemi, M., P. Dehghanian, Y. Darestani, and J. Su. 2023. "Parameterized Wildfire Fragility Functions for Overhead Power Line Conductors." *IEEE Transactions on Power Systems* 39, no. 2: 2517–2527. https://doi.org/10.1109/TPWRS.2023.3298769.

NERC, N. A. R. C. 2022. State of Reliability—An Assessment of 2021 Bulk Power System Performance. https://www.nerc.com/pa/RAPA/PA/Performance%20Analysis%20DL/NERC_SOR_2022.pdf.

News, S. 2021. Germany and Belgium Floods: More Than 50 Dead and Over 70 Missing After Heavy Rain. https://web.archive.org/web/20210715200906/https://news.sky.com/story/germany-and-belgium-floods-at-least-44-dead-and-more-than-70-missing-after-heavy-rain-12356134.

Nicolas, C., J. Rentschler, A. P. van Loon, et al. 2019. "Stronger Power: Improving Power Sector Resilience to Natural Hazards." In *Sector Note for LIFELINES: The Resilient Infrastructure Opportunity*. Washington, DC: World Bank.

Nirandjan, S., E. E. Koks, M. Ye, et al. 2024. "Review Article: Physical Vulnerability Database for Critical Infrastructure Multi-Hazard Risk Assessments—A Systematic Review and Data Collection." *Natural Hazards and Earth System Sciences Discussions*. https://doi.org/10.5194/nhess-2023-208.

Noebels, M., R. Preece, and M. Panteli. 2022. "A Machine Learning Approach for Real-Time Selection of Preventive Actions Improving Power Network Resilience." *IET Generation, Transmission and Distribution* 16, no. 1: 181–192. https://doi.org/10.1049/gtd2.12287.

NYT. 2021. European Floods are Latest Sign of a Global Warming Crisis. https://www.nytimes.com/2021/07/16/world/europe/germany-floods-climate-change.html.

Omogoye, S. O., K. A. Folly, and K. O. Awodele. 2023. "A Comparative Study Between Bayesian Network and Hybrid Statistical Predictive Models for Proactive Power System Network Resilience Enhancement Operational Planning." *IEEE Access* 11: 41281–41302. https://doi.org/10.1109/ACCESS.2023.3263490.

Onyewuchi, U. P., A. Shafieezadeh, M. M. Begovicieee, and R. Desroches. 2015. "A Probabilistic Framework for Prioritizing Wood Pole Inspections Given Pole Geospatial Data." *IEEE Transactions on Smart Grid* 6: 973–979. https://doi.org/10.1109/TSG.2015.2391183.

Palin, E. J., I. Stipanovic Oslakovic, K. Gavin, and A. Quinn. 2021. "Implications of Climate Change for Railway Infrastructure." *WIREs Climate Change* 12, no. 5: 1–41. https://doi.org/10.1002/wcc.728.

Pan, H., C. Li, and L. Tian. 2021. "Seismic Fragility Analysis of Transmission Towers Considering Effects of Soil-Structure Interaction and Depth-Varying Ground Motion Inputs." *Bulletin of Earthquake Engineering* 19: 4311–4337. https://doi.org/10.1007/s10518-021-01124-x.

Panteli, M., and P. Mancarella. 2015. "Influence of Extreme Weather and Climate Change on the Resilience of Power Systems: Impacts and Possible Mitigation Strategies." *Electric Power Systems Research* 127: 259–270. https://doi.org/10.1016/j.epsr.2015.06.012.

Panteli, M., and P. Mancarella. 2017. "Modeling and Evaluating the Resilience of Critical Electrical Power Infrastructure to Extreme Weather Events." *IEEE Systems Journal* 11, no. 3: 1733–1742. https://doi.org/10.1109/JSYST.2015.2389272.

Panteli, M., C. Pickering, S. Wilkinson, R. Dawson, and P. Mancarella. 2017. "Power System Resilience to Extreme Weather: Fragility Modeling, Probabilistic Impact Assessment, and Adaptation Measures." *IEEE Transactions on Power Systems* 32, no. 5: 3747–3757. https://doi.org/10.1109/TPWRS.2016.2641463.

PG&E. 2021. Technosylva 2019 PSPS Event Wildfire Risk Analysis Reports. https://www.cpuc.ca.gov/-/media/cpuc-website/divisions/safety-and-enforcement-division/documents/technosylva-report-on-pge-psps-eventoct-912-2019.pdf.

Pöyhönen, J. 2022. "Cyber Security of an Electric Power System in Critical Infrastructure." In *Cyber Security. Computational Methods in Applied Sciences*, edited by M. Lehto and P. Neittaanmäki, vol. 56, 217–239. Cham: Springer. https://doi.org/10.1007/978-3-030-91293-2_9.

Prasad, G. G., and S. Banerjee. 2013. "The Impact of Flood-Induced Scour on Seismic Fragility Characteristics of Bridges." *Journal of Earthquake Engineering* 17, no. 6: 803–828. https://doi.org/10.1080/13632469.2013.771593.

Raj, S. V., U. Bhatia, and M. Kumar. 2022. "Cyclone Preparedness Strategies for Regional Power Transmission Systems in Data-Scarce Coastal Regions of India." *International Journal of Disaster Risk Reduction* 75: 102957. https://doi.org/10.1016/j.ijdrr.2022.102957.

Reed, D. A., C. J. Friedland, S. Wang, and C. C. Massarra. 2016. "Multi-Hazard System-Level Logit Fragility Functions." *Engineering Structures* 122: 14–23. https://doi.org/10.1016/j.engstruct.2016.05.006.

Reed, D. A., M. D. Powell, and J. M. Westerman. 2010. "Energy Infrastructure Damage Analysis for Hurricane Rita." *Natural Hazards Review* 11: 102–109. https://doi.org/10.1061/(asce)nh.1527-6996.0000012.

Reed, D., S. Wang, K. Kapur, and C. Zheng. 2016. "Systems-Based Approach to Interdependent Electric Power Delivery and Telecommunications Infrastructure Resilience Subject to Weather-Related Hazards." *Journal of Structural Engineering* 142, no. 8: C4015011. https://doi.org/10.1061/(asce)st.1943-541x.0001395.

Reed, D. A., and S. Wang. 2018. "Numerical Modeling of Power Delivery and Telecommunications Infrastructure for Hurricanes Harvey and Irma." In *Forensic Engineering 2018: Forging Forensic Frontiers—Proceedings of the 8th Congress on Forensic Engineering*. Reston, VA. https://doi.org/10.1061/9780784482018.097.

ReFLOAT-ER. 2023. "*Project on Resilient Reconstruction of Flooded Appenines Territories of Emilia-Romagna*." Personal Communication During the Mission in Modigliana Municipality. https://www.dabc.polimi.it/wp-content/uploads/2023/12/resilienza.pdf.

Rezaei, S. N., L. Chouinard, S. Langlois, and F. Légeron. 2017. "A Probabilistic Framework Based on Statistical Learning Theory for Structural Reliability Analysis of Transmission Line Systems." *Structure and Infrastructure Engineering* 13: 1538–1552. https://doi.org/10.1080/15732479.2017.1299771.

Roege, P. E., Z. A. Collier, J. Mancillas, J. A. McDonagh, and I. Linkov. 2014. "Metrics for Energy Resilience." *Energy Policy* 72: 249–256. https://doi.org/10.1016/j.enpol.2014.04.012.

RTO. 2021. Wildfires Raise Concerns for Western Tx Lines. https://www.rtoinsider.com/28328-wildfires-raise-concerns-for-western-tx-lines/.

Sahraei-Ardakani, M., and G. Ou. 2018. "Day-Ahead Preventive Scheduling of Power Systems During Natuaral Hazards via Stochastic Optimization." *IEEE Power and Energy Society General Meeting*: 1. https://doi.org/10.1109/PESGM.2017.8274453.

Salman, A. M., and Y. Li. 2016. "Age-Dependent Fragility and Life-Cycle Cost Analysis of Wood and Steel Power Distribution Poles Subjected to Hurricanes." *Structure and Infrastructure Engineering* 12: 890–903. https://doi.org/10.1080/15732479.2015.1053949.

Salman, A. M., and Y. Li. 2017. "Assessing Climate Change Impact on System Reliability of Power Distribution Systems Subjected to Hurricanes." *Journal of Infrastructure Systems* 23, no. 1: 04016024. https://doi.org/10.1061/(asce)is.1943-555x.0000316.

Sánchez-Muñoz, D., J. L. Domínguez-García, E. Martínez-Gomariz, B. Russo, J. Stevens, and M. Pardo. 2020. "Electrical Grid Risk Assessment Against Flooding in Barcelona and Bristol Cities." *Sustainability (Switzerland)* 12, no. 4: 1527. https://doi.org/10.3390/su12041527.

Sayers, P., M. Horritt, S. Carr, et al. 2020. *Third UK Climate Change Risk Assessment (CCRA3): Future Flood Risk*. Main report - Final report prepared for the Committee on Climate Change, London, UK: Committee on Climate Change, 102 pp. (UKCEH Project no. C07063).

Scherb, A., L. Garrè, and D. Straub. 2019. "Evaluating Component Importance and Reliability of Power Transmission Networks Subject to Windstorms: Methodology and Application to the Nordic Grid." *Reliability Engineering and System Safety* 191: 106517. https://doi.org/10.1016/j.ress.2019.106517.

Schultz, M. T., B. P. D. Gouldby, J. Simm, and L. J. Wibowo. 2010. ERDC SR-10-1 Beyond the Factor of Safety: Developing Fragility Curves to Characterize System Reliability.

Schweikert, A., L. Nield, E. Otto, and M. Deinert. 2019. *Resilience and Critical Power System Infrastructure; Lessons Learned From Natural Disasters and Future Research Needs* (Policy Research Working Paper; No. 8900). Washington, DC: World Bank. https://openknowledge.worldbank.org/entities/publication/47095572-0e4e-5c0b-a591-69c91e1e4812.

Schweikert, A. E., and M. R. Deinert. 2021. "Vulnerability and Resilience of Power Systems Infrastructure to Natural Hazards and Climate Change." *WIREs Climate Change* 12, no. 5: e724. https://doi.org/10.1002/wcc.724.

Serrano, R., M. Panteli, and A. Parisio. 2023. "Risk Assessment of Power Systems Against Wildfires." In *IEEE Belgrade PowerTech*. Belgrade, Serbia: PowerTech. https://doi.org/10.1109/PowerTech55446.2023.10202967.

Serrano-Fontova, A., H. Li, Z. Liao, et al. 2023. "A Comprehensive Review and Comparison of the Fragility Curves Used for Resilience Assessments in Power Systems." *IEEE Access* 11: 108050–108067. https://doi.org/10.1109/ACCESS.2023.3320579.

Shafieezadeh, A., P. U. Onyewuchi, M. M. Begovic, and R. DesRoches. 2013. "Fragility Assessment of Wood Poles in Power Distribution Networks Against Extreme Wind Hazards." In *Advances in Hurricane Engineering: Learning From Our Past—Proceedings of the 2012 ATC and SEI Conference on Advances in Hurricane Engineering*. Resto, VA: American Society of Civil engineers (ASCE). https://doi.org/10.1061/9780784412626.074.

Shafieezadeh, A., U. P. Onyewuchi, M. M. Begovic, and R. Desroches. 2014. "Age-Dependent Fragility Models of Utility Wood Poles in Power Distribution Networks Against Extreme Wind Hazards." *IEEE Transactions on Power Delivery* 29: 131–139. https://doi.org/10.1109/TPWRD.2013.2281265.

Shaoyun, G., L. Jifeng, L. Hong, C. Yuchen, Y. Zan, and Y. Jun. 2019. "Assessing and Boosting the Resilience of a Distribution System Under Extreme Weather." In *IEEE Power and Energy Society General Meeting (PESGM)*, Atlanta, GA, 1–5. https://doi.org/10.1109/PESGM40551.2019.8974012.

Shehata, A. Y., A. A. El Damatty, and E. Savory. 2005. "Finite Element Modeling of Transmission Line Under Downburst Wind Loading." *Finite Elements in Analysis and Design* 42, no. 1: 71–89. https://doi.org/10.1016/j.finel.2005.05.005.

Silva, V., S. Akkar, J. Baker, et al. 2019. "Current Challenges and Future Trends in Analytical Fragility and Vulnerability Modeling." *Earthquake Spectra* 35, no. 4: 1927–1952. https://doi.org/10.1193/042418EQS101O.

Snaiki, R., and S. S. Parida. 2023. "A Data-Driven Physics-Informed Stochastic Framework for Hurricane-Induced Risk Estimation of Transmission Tower-Line Systems Under a Changing Climate." *Engineering Structures* 280: 115673. https://doi.org/10.1016/j.engstruct.2023.115673.

Solheim, O. R., and T. Trotscher. 2018. "Modelling Transmission Line Failures Due to Lightning Using Reanalysis Data and a Bayesian Updating Scheme." In *International Conference on Probabilistic Methods Applied to Power Systems, PMAPS 2018—Proceedings*, Boise, ID. https://doi.org/10.1109/PMAPS.2018.8440365.

Sonal, and D. Ghosh. 2021. "Fuzzy Logic-Based Planning and Operation of a Resilient Microgrid." In *Applications of Fuzzy Logic in Planning and Operation of Smart Grids, Power Systems*, edited by M. Rahmani-Andebili. Cham: Springer. https://doi.org/10.1007/978-3-030-64627-1_2.

Souto, L., P. C. Taylor, and J. Wilkinson. 2022. "Probabilistic Impact Assessment of Lightning Strikes on Power Systems Incorporating Lightning Protection Design and Asset Condition." *International Journal of Electrical Power & Energy Systems* 148: 108974. https://doi.org/10.1016/j.ijepes.2023.108974.

Sovacool, B. K. 2011. "Evaluating Energy Security in the Asia Pacific: Towards a More Comprehensive Approach." *Energy Policy* 39, no. 11: 7472–7479. https://doi.org/10.1016/j.enpol.2010.10.008.

Sperstad, I. B., G. H. Kjølle, and O. Gjerde. 2020. "A Comprehensive Framework for Vulnerability Analysis of Extraordinary Events in Power Systems." *Reliability Engineering & System Safety* 196: 106788. https://doi.org/10.1016/j.ress.2019.106788.

Spiegel. 2005. The German Power Outage—Brittle Giants. https://www.spiegel.de/international/spiegel/the-german-power-outage-brittle-giants-a-388759.html.

Tavares da Costa, R., and E. Krausmann. 2021. Impacts of Natural Hazards and Climate Change on EU Security and Defence. https://doi.org/10.2760/244397.

Tavares da Costa, R., E. Krausmann, and C. Hadjisavvas. 2023. *Impacts of Climate Change on Defence-Related Critical Energy Infrastructure* (Issue June). Luxembourg: Publications Office of the European Union. https://doi.org/10.2760/03454.

Teoh, Y. E., A. Alipour, and A. Cancelli. 2019. "Probabilistic Performance Assessment of Power Distribution Infrastructure Under Wind Events." *Engineering Structures* 197: 109199. https://doi.org/10.1016/j.engstruct.2019.05.041.

Tian, L., X. Zhang, and X. Fu. 2020. "Fragility Analysis of a Long-Span Transmission Tower-Line System Under Wind Loads." *Advances in Structural Engineering* 23: 2110–2120. https://doi.org/10.1177/1369433220903983.

Tibolt, M., M. Bezas, I. Vayas, and J. Jaspart. 2021. "The Design of a Steel Lattice Transmission Tower in Central Europe." *Ce/Papers* 4, no. 2–4: 243–248. https://doi.org/10.1002/cepa.1288.

Turner, S. W. D., N. Voisin, J. Fazio, D. Hua, and M. Jourabchi. 2019. "Compound Climate Events Transform Electrical Power Shortfall Risk in the Pacific Northwest." *Nature Communications* 10: 8. https://doi.org/10.1038/s41467-018-07894-4.

UNDRR. 2015. *United Nations Office for Disaster Risk Reduction: Sendai Framework for Disaster Risk Reduction 2015–2030*, 37. Geneva, Switzerland: United Nations Office for Disaster Risk Reduction. https://www.undrr.org/media/16176/download?startDownload=20240525.

UNDRR. 2023. Principles for Resilient Infrastructure. https://www.undrr.org/media/78694/download?startDownload=20240525.

Van Vliet, M. T. H., J. R. Yearsley, F. Ludwig, S. Vögele, D. P. Lettenmaier, and P. Kabat. 2012. "Vulnerability of US and European Electricity Supply to Climate Change." *Nature Climate Change* 2: 676–681. https://doi.org/10.1038/nclimate1546.

Venkatasubramanian, B. V., M. Lotfi, P. Mancarella, et al. 2023. "Machine Learning Based Identification and Mitigation of Vulnerabilities in Distribution Systems Against Natural Hazards." In *27th International Conference on Electricity Distribution (CIRED 2023)*, 2908–2912, Stevenage, UK: Institution of Engineering and Technology (IET). https://doi.org/10.1049/icp.2023.0985.

VOSviewer. 2023. 1.6.19 Software. https://www.vosviewer.com/download.

Wahlin, B., C. Davis, and P. Kandaris. 2011. "Scour Evaluation for Protection of Critical Infrastructure on the Hassayampa River." *World Environmental and Water Resources Congress* 2011: 2749–2758. https://doi.org/10.1061/41173(414)286.

Wang, J., H. N. Li, X. Fu, Z. Q. Dong, and Z. G. Sun. 2022. "Wind Fragility Assessment and Sensitivity Analysis for a Transmission Tower-Line System." *Journal of Wind Engineering and Industrial Aerodynamics* 231: 105233. https://doi.org/10.1016/j.jweia.2022.105233.

Wang, Z., and Z. Wang. 2023. "A Novel Preventive Islanding Scheme of Power System Under Extreme Typhoon Events." *International Journal of Electrical Power & Energy Systems* 147: 108857. https://doi.org/10.1016/j.ijepes.2022.108857.

Watson, P. L., W. Hughes, D. Cerrai, W. Zhang, A. Bagtzoglou, and E. Anagnostou. 2024. "Integrating Structural Vulnerability Analysis and Data-Driven Machine Learning to Evaluate Storm Impacts on the

Power Grid." *IEEE Access* 12: 63568–63583. https://doi.org/10.1109/ACCESS.2024.3396414.

Winkler, J., L. Dueñas-Osorio, R. Stein, and D. Subramanian. 2010. "Performance Assessment of Topologically Diverse Power Systems Subjected to Hurricane Events." *Reliability Engineering and System Safety* 95, no. 4: 323–336. https://doi.org/10.1016/j.ress.2009.11.002.

Xue, J., F. Mohammadi, X. Li, M. Sahraei-Ardakani, G. Ou, and Z. Pu. 2020. "Impact of Transmission Tower-Line Interaction to the Bulk Power System During Hurricane." *Reliability Engineering and System Safety* 203: 107079. https://doi.org/10.1016/j.ress.2020.107079.

Xue, J., Z. Xiang, and G. Ou. 2021. "Predicting Single Freestanding Transmission Tower Time History Response During Complex Wind Input Through a Convolutional Neural Network Based Surrogate Model." *Engineering Structures* 233: 111859. https://doi.org/10.1016/j.engstruct.2021.111859.

Yuan, H., W. Zhang, J. Zhu, and A. C. Bagtzoglou. 2018. "Resilience Assessment of Overhead Power Distribution Systems Under Strong Winds for Hardening Prioritization." *ASCE-ASME Journal of Risk and Uncertainty in Engineering Systems, Part A: Civil Engineering* 4, no. 4: 04018037. https://doi.org/10.1061/AJRUA6.0000988.

Zhang, C., C. Song, and A. Shafieezadeh. 2022. "Adaptive Reliability Analysis for Multi-Fidelity Models Using a Collective Learning Strategy." *Structural Safety* 94: 102141. https://doi.org/10.1016/j.strusafe.2021.102141.

Zhang, J., Y. Bagtzoglou, J. Zhu, B. Li, and W. Zhang. 2023. "Fragility-Based System Performance Assessment of Critical Power Infrastructure." *Reliability Engineering and System Safety* 232: 109065. https://doi.org/10.1016/j.ress.2022.109065.

Zhaohong, B., L. Yanlng, L. Gengfeng, and L. Furong. 2017. "Battling the Extreme: A Study on the Power System Resilience." *Proceedings of the IEEE* 105, no. 7: 1–14. https://doi.org/10.1109/JPROC.2017.2679040.

**Appendix A**

Some references reported in Table A1 are referred to both as analytical and empirical, because they include fragility models that fall into both categories.

TABLE A1 | The fragility models for power grid assets in the last two decades.

| Asset | Hazard | Region | Fragility model | IMs | Publication year |
|---|---|---|---|---|---|
| Transmission towers and/or lines | Wind | EU [1]<br>US [2], [3], [4], [5], [6], [7], [8], [9], [10], [11], [12], [13]<br>Asia [14], [15], [16], [17], [18], [19], [20], [21], [22], [23], [24], [25], [26] | Analytical: [2], [14], [5], [15], [16], [1], [6], [17], [7], [20], [8], [21], [9], [10], [11], [23], [24], [25], [26]<br>Empirical: [3], [4], [18], [19], [22], [12], [13] | Maximum sustained wind speed: [5], [2], [19], [8], [21], [9], [22], [23], [24]<br>Peak wind gust: [3]<br>Wind speed ratio squared: [4]<br>Basic wind speed: [14], [15], [17], [18]<br>10-min average wind speed: [16], [25], [26]<br>3-s gust wind speed: [1], [7], [20], [12]<br>Mean wind speed: [6], [10], [11], [13] | 2023: [6], [18]<br>2022: [3], [15], [20], [9], [23], [24], [25]<br>2021: [5], [11], [26]<br>2020: [2], [7], [19], [21], [10]<br>2019: [14], [16], [1]<br>2018: [17], [8], [22], [12]<br>2010: [4], [13] |
| | Heavy snow and lightning | EU [27]<br>Asia [28], [29] | Empirical [28], [27], [29] | Ice accretion: [28]<br>Lightning indices (K, TT): [27]<br>Thunderstorm duration: [29] | 2018: [28]<br>Lightning indices (K, TT): [27] |
| | Multi-hazard | US [30], [31]<br>Asia [32], [33], [34], [35] | Analytical: [32], [33], [35], [31], [34]<br>Empirical: [30] | Wind speed and rain rate: [32], [33], [31], [34]<br>PGA & wind speed: [30], [35] | 2023: [31], [34]<br>2022: [35]<br>2020: [33]<br>2018: [30]<br>2016: [32] |
| Distribution poles and lines | Wind | EU [36]<br>US [37], [38], [39], [40], [41], [42], [43], [44], [45], [46], [47], [48]<br>Asia [49], [50], [51], [52], [53]<br>Africa [54] | Empirical [36], [49], [50], [54], [53]<br>Analytical [37], [38], [39], [40], [41], [42], [43], [44], [45], [46], [51], [52], [47], [48] | Maximum wind speed: [36], [49], [40], [43], [44], [50], [54], [51], [52], [53]<br>Maximum 3 s wind gust: [37], [39]<br>3-s gust wind speed: [38], [42], [45], [47], [48]<br>Basic wind speed: [41], [46] | 2023: [40], [54], [51], [47]<br>2022: [43]<br>2021: [49], [50], [53]<br>2020: [41]<br>2019: [52]<br>2018: [36], [48]<br>2017: [44]<br>2015: [37], [38]<br>2014: [39]<br>2013: [45], [46] |
| | Multi-hazard | US [55], [56], [57], [58], [59] | Empirical: [55], [56]<br>Analytical: [57], [59], [58] | Maximum sustained wind speed, rainfall and storm surge: [55]<br>Maximum sustained wind speed and storm surge: [56]<br>Maximum wind speed and radial ice thickness: [57]<br>3-s gust wind speed and ice thickness & tree weight and ice thickness: [58]<br>Mean wind speed and radial ice thickness: [59] | 2023: [58]<br>2020: [57]<br>2019: [59]<br>2016: [55], [56] |

(Continues)

**TABLE A1** | (Continued)

| Asset | Hazard | Region | Fragility model | IMs | Publication year |
|---|---|---|---|---|---|
| Substations | Flood | US [60]<br>Asia [61] | Empirical: [60]<br>Analytica: [61] | Water depth: [60], [61] | 2013: [60]<br>2024: [61] |

*Note:* [1] Scherb, Garrè, and Straub (2019), [2] Ma, Bocchini, and Christou (2020) and Ma, Dai, and Pang (2020), [3] Dos Reis et al. (2022), [4] Reed, Powell, and Westerman (2010), [5] Ma, Khazaali, and Bocchini (2021), [6] Alipour and Dikshit (2023), [7] Hou and Chen (2020), [8] Sahraei-Ardakani and Ou (2018), [9] Ma, Christou, and Bocchini (2022), [10] Xue et al. (2020), [11] Xue, Xiang, and Ou (2021), [12] Reed and Wang (2018), [13] Winkler et al. (2010), [14] Fu et al. (2019), [15] Liu and Yan (2022), [16] Cai et al. (2019), [17] Fu and Li (2018), [18] Wang and Wang (2023), [19] Liu et al. (2020), [20] Raj, Bhatia, and Kumar (2022), [21] Tian, Zhang, and Fu (2020), [22] Liu, Lei, and Hou (2019), [23] Chen et al. (2022), [24] Li, Jia, et al. (2022), [25] Wang et al. (2022), [26] Cai and Wan (2021), [27] Solheim and Trotscher (2018), [28] Lu et al. (2018), [29] Bao et al. (2021), [30] Rezaei et al. (2017), [31] Snaiki and Parida (2023), [32] Fu, Li, and Li (2016), [33] Fu et al. (2020), [34] Bi et al. (2023), [35] Li, Pan, et al. (2022), [36] Dunn et al. (2018), [37] Onyewuchi et al. (2015), [38] Salman and Li (2016), [39] Shafieezadeh et al. (2014), [40] Zhang et al. (2023), [41] Lee and Ham (2021), [42] Darestani and Shafieezadeh (2019), [43] Lu and Zhang (2022), [44] Salman and Li (2017), [45] Bjarnadottir, Li, and Stewart (2013), [46] Shafieezadeh et al. (2013), [47] Hou and Muraleetharan (2023), [48] Yuan et al. (2018), [49] Najafi Tari, Sepasian, and Tourandaz Kenari (2021), [50] Karimi, Ravadanegh, and Haghifam (2021), [51] Dehghani, Mohammadi, and Karimi (2023), [52] Shaoyun et al. (2019), [53] Sonal and Ghosh (2021), [54] Omogoye, Folly, and Awodele (2023), [55] Reed, Friedland, et al. (2016), [56] Wang, et al. (2016), [57] Ma, Bocchini, and Christou (2020) and Ma, Dai, and Pang (2020), [58] Hou et al. (2023), [59] Teoh, Alipour, and Cancelli (2019), [60] FEMA (2013), [61] Li et al. (2024).